\documentclass[a4paper,11pt]{article}
\usepackage{jheppub} 
\usepackage{appendix}
\usepackage{graphicx}
\usepackage{physics}
\usepackage{tensor}
\usepackage{todonotes}
\usepackage{amsthm}
\usepackage{amsmath}
\usepackage{lmodern}
\usepackage{eso-pic}
\usepackage{transparent}
\usepackage{graphicx}
\usepackage{graphicx}
\usepackage{dcolumn}
\usepackage{bm}
\usepackage{hyperref}
\usepackage[mathlines]{lineno}
\usepackage[dvipsnames]{xcolor}

\newcommand{\hillingas}{NC gas}

\usepackage{setspace}
\usepackage{etoolbox}

\usepackage{lmodern}
\usepackage{eso-pic}
\usepackage{transparent}
\usepackage{graphicx}

\begin{document}{}

\title{On Black Holes Surrounded by Radiation
II: Thermodynamics}

\author{Marcos Riojas \& Matthew J. Strassler}
\affiliation{Center for the Fundamental Laws of Nature, Harvard University, Cambridge, MA, USA}
\emailAdd{marcos\_riojas@fas.harvard.edu}
\emailAdd{strassler@g.harvard.edu}

\abstract{
In a companion paper we considered a Schwarzschild black hole of mass $m$ enveloped by a thick ``ocean'' of massless particles that extends the black hole's photon sphere into a region of finite depth. There we showed that this ``hillingar black hole'', of ADM mass $M$, optically mimics an ordinary black hole of the same mass. Here we find it also mimics the black hole thermodynamically: the formal assumption of thermal equilibrium  implies the system has the same temperature and entropy as an ordinary black hole of mass $M$.  
We check this result carefully using multiple methods; a further method and indications of metastability are given by one of us in a companion paper. In AdS space, the mimicry does not hold, and the solutions have a richer structure. While it is far from clear that these systems are models for more realistic ones, we note possible connections with black hole evolution. In particular, assuming thermal equilibrium can be established and maintained, an HBH in a cavity of radius $\geq 3M$ can evaporate, potentially posing the information puzzle in a small finite volume.}

\maketitle
\flushbottom

\section{\label{sec:level1}Introduction\protect}

In a previous paper (Paper I) we examined static spherically symmetric solutions of mass $M$ in which a Schwarzschild black hole of mass $m<M$ is surrounded by a self-gravitating gas \cite{Riojas:2026ytt}.  Massless radiation orbiting the black hole on randomly oriented circular orbits was of particular interest. This null gas, with zero radial pressure, is an ultra-relativistic Einstein cluster; we will refer to it as a ``null cluster'' or NC gas, and to a  black hole surrounded by a thick shell or ``ocean'' of such gas  as a ``hillingar black hole'' (HBH).\footnote{``Hillingar'' is an Old Norse term for an arctic mirage that upheaves images and changes the apparent position of the horizon. The mirage can cause light rays to curve at the same rate as the water's surface, making it appear flat;  the ocean of an HBH, which sits on an extended photon sphere, similarly appears flat to an observer within it. }  Among other features, an HBH has a unique and extreme form, lies entirely behind its photon sphere, and optically imitates a black hole as seen from observers outside its photon sphere. In particular it has the same shadow and the same single photon ring as a black hole of the same total mass. We also suggested that such an ``ocean''-like shell might be found in nature, at least briefly.\footnote{For instance, after a collision of two black holes, one might imagine that gravitational waves captured in orbit of the post-collision black hole could be sufficiently abundant to backreact and significantly distort the metric, perhaps affecting the black hole's ringdown phase.}

In this paper we continue our study of these objects by exploring their possible thermodynamic properties, assuming equilibrium between black hole and ocean. This assumption is formal --- no such equilibrium may  be possible --- and we should therefore view this system as a toy model, one that might describe a more realistic system of a black hole and surrounding radiation. 
But with this important caveat in mind, we find that an HBH in asymptotically flat space has temperature, free energy and entropy identical to those of an ordinary black hole of precisely the same ADM mass. These properties are unique to the HBH in various senses, as shown below and in a companion paper \cite{Riojas:2026god}.

\begin{figure}
    \centering
    \includegraphics[width=\linewidth]{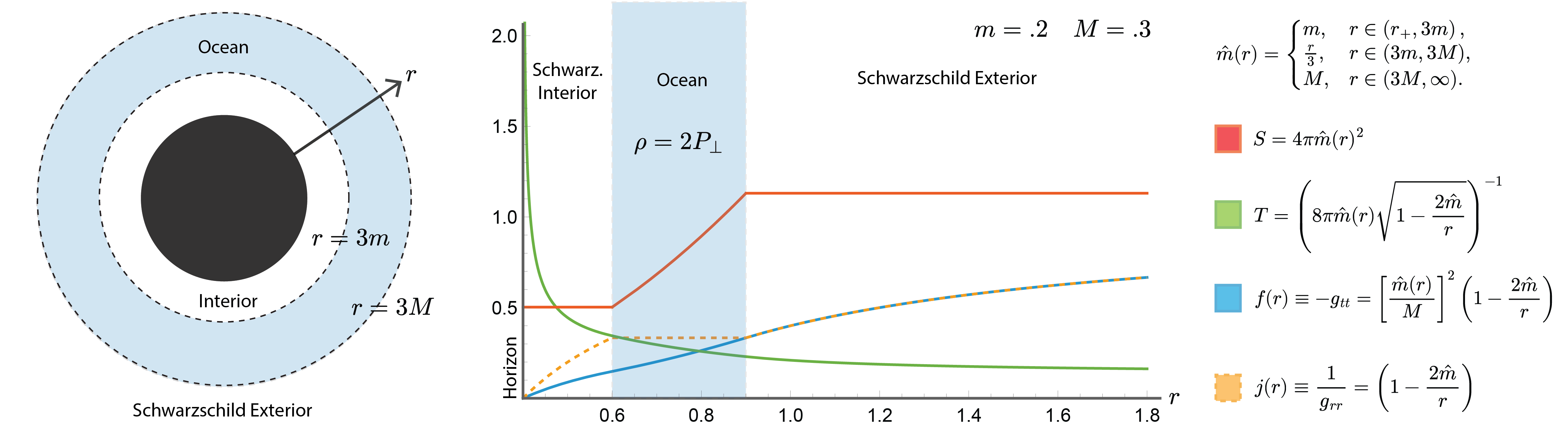}
    \caption{Hillingar black holes consist of a black hole surrounded by an ``ocean" of orbiting massless particles. The continuous mass function $\widehat{m}(r)$ is piecewise defined, taking the value $m$ within the Schwarzschild interior, $M$ within the Schwarzschild exterior, and $\widehat{m}=r/3$ within the ocean. This in turn affects the metric functions $f(r)$ and $j(r)$, as well as the local value of the entropy $S(r)$ and temperature $T(r)$, as defined in York \cite{York:1986it}. In the limit where $\rho \rightarrow 0$, the ocean becomes a regular photon sphere and the Schwarzschild metric is recovered.
    }
    \label{fig:Metric}
\end{figure}

We first outline our main results.
As sketched in Fig.~\ref{fig:Metric}, we consider (in asymptotically flat space) an HBH  of total mass $M$, consisting of a central black hole of mass $m$ surrounded by a  ocean of NC gas. 
The HBH has the metric 
\begin{eqnarray}\label{fjmetric0}
    d s^2&=&-f(r) d t^2+j(r)^{-1} d r^2+r^2 d \theta^2+r^2 \sin ^2(\theta) d \phi^2 \\[4pt]
    j(r)&=&  1
    -\frac{2\widehat{m}(r)}{r}  \quad , \quad \quad 
    f(r) =\left[\frac{{\widehat{m}}(r)}{M}\right]^2  j(r) \ .
\end{eqnarray}
The mass function  ${\widehat{m}}(r)$, the mass interior to a sphere of areal radius $r$, takes the piecewise form  
\begin{equation}\label{massfunction0}
\widehat{m}(r)=
\begin{cases}
m, & r\in(r_+,\,3m),\\[2pt]
\dfrac{r}{3}, & r\in(3m,\,3M),\\[2pt]
M, & r\in(3M,\,\infty).
\end{cases}
\end{equation}
This
grows linearly from the black hole's photon sphere at $r=3m$ to the HBH system's photon sphere at $r=3M$, where it reaches the HBHs ADM mass $M$. 
Similar solutions exist in all higher dimensions, as shown in Paper I \cite{Riojas:2026ytt} and reviewed in Appendix \ref{app:otherHBH}.

Although the HBH is simple, it seems to have been little-studied.  Indeed we have only found the metric in \cite{DiFilippo:2024poc}, where it was argued to be unstable, and as a special case of Model I in \cite{Maeda:2024tsg}, where its point-by-point marginal stability was noted. The latter in turn generalizes \cite{Cardoso:2021wlq}, where Einstein clusters are extended to include horizons; see also \cite{Jusufi:2022jxu}.  In an even earlier study \cite{Andreasson:2021lsh} a discrete version of the HBH ocean, made of closely spaced shells of finite thickness comprised of massless particles, is discussed.

To put the system in formal thermal equilibrium, we first obtain the Hawking temperature measured at infinity, $T_H$, from the surface gravity at the central black hole's horizon.  Although the black hole would have $T_H=1/8\pi Gm$ in the absence of the ocean, we find in Sec.~\ref{sec:thermo} that the HBH has temperature
\begin{equation}
    T_{H}=\frac{1}{8\pi G M} \ .
\end{equation}
This remarkable $m$-independent result indeed matches the Hawking temperature of a black hole of mass $M$.

Since the HBH and ordinary black hole share a mass and a temperature, one might already wonder if their entropies\footnote{Throughout this paper, ``entropy'' generally refers to ``coarse-grained entropy.'' Our methods are unsuitable for a direct calculation of fine-grained entropy.} match. In Secs.~\ref{sec:powers}-\ref{sec:nomatteraction}, we demonstrate this is true using two approaches, one based on the Euclidean action and one on direct entropy maximization. In a companion paper \cite{Riojas:2026god}, a third method computes thermodynamic quantities by building up layers of self-gravitating shells and matter, further confirming our results.

The result from the  Euclidean action is the most immediate, so we sketch it here.  A black hole's action is given just by the Gibbons-Hawking/York boundary term \cite{Gibbons:1976ue,York:1986it}, but in most systems with matter we would expect a bulk gravity term and a matter action term.  However, the NC gas has $T^{\ \mu}_\mu\propto R^{\ \mu}_\mu=0$ everywhere, eliminating the bulk term, and we will see in Sec.~\ref{subsec:matteractionzero} that its matter action vanishes as well.  Since only the GHY boundary term remains, which depends on the ADM mass $M$ but not on $m$, the HBH calculation is the same as that of a black hole.  
This holds even when $M\gg m$, thus implying that a large ball of \hillingas\ around a tiny black hole has an enormous entropy of order $GM^2/\hbar$.  
For the NC gas to carry comparable entropy to a black hole requires an entropy density of the form
\begin{equation}
    s(r) = \frac{\rho(r)}{T(r)} \propto \frac{1}{\hbar Gr}
\end{equation}
which is extreme, though smaller than Planckian density.

In addition, \cite{Riojas:2026god} shows that the standard obstruction to equilibrium between self-gravitating matter and an asymptotically-flat Schwarzschild black hole uniquely selects the HBH. Thin shells around a Schwarzschild black hole are mechanically stable only outside the photon sphere \cite{Brady:1991np}, while the local heat capacity of Hawking radiation is positive only inside it \cite{York:1986it}. The HBH straddles the boundary, as a continuum of orbiting particles extending the photon sphere into a region of finite depth.

Corresponding results hold in all dimensions $d\geq 4$.  In AdS spaces, the entropy and temperature now depend on $m$ as well as $M$, as we will describe in detail in a future paper (Paper III) \cite{RiojasStrasslerAdS}; we will briefly mention some key findings and their potential implications in Secs.~\ref{sec:AdS}-\ref{sec:cavity}. Results that might suggest a possible connection with black hole evaporation, and discussion thereof, are found in Secs.~\ref{sec:cavity}-\ref{sec:meanings}.

\subsection{Important Caveats}

All of the results just listed rest on the strong assumption that an NC gas can be in quasi-equilibrium with a black hole. But there are many reasons to doubt this is possible.  As we mentioned above, our view is that {\it one should regard the equilibrium-HBH system as a toy model for a realistic system that might or might not exist.} The Hawking temperature $T_H\sim m_{pl}^2/M$ (where $m_{pl}$ is the Planck mass) is so low that modes in the ocean with energy $E_\gamma\sim T_H$ have wavelength comparable to the ocean's radius $r\sim \ell_{pl}(M/m_{pl})$, where $\ell_{pl}$ is the Planck length.  In quantum field theory, we would not expect such long-wavelength modes could be placed into a statistically averaged state whose  $T_\mu^\nu$ has sharp edges at $r=3m$ and $3M$, as in Eqs.~\eqref{fjmetric0}-\eqref{massfunction0}. Other concerns include the possible tendency of particles to spiral off the ocean's surfaces, the unclear nature of equilibrium in a  system with zero radial pressure, and the enormous entropy density $s \sim 1/Gr$.  Faced with these issues, one cannot be satisfied with simply treating the NC gas using its energy-momentum tensor, and should instead seek a microscopic description of it in a realistic theory.  

As with all toy models, the  question of what lessons we should learn from it cannot  be answered within the model itself.  
However, because the model has many unique and interesting features, among them thermodynamic and optical mimicry, one may wonder if it touches on something significant, at least at the formal level, and perhaps even in the context of deeper questions about black hole evolution. We will therefore calculate its properties without prejudice, reserving judgment until Sec.~\ref{sec:meanings}, where we consider the model's strengths and weaknesses and its possible connections with the structure of general relativity and with black hole evolution.

\subsection{Past Literature}

Our work is related to numerous other studies in the literature.  
Since black holes with orbiting matter are a current observational target, studies of their properties form a vast literature. For classical properties of these systems, see Paper I \cite{Riojas:2026ytt}.

The thermodynamics of self-gravitating gases has been widely studied.  The isotropic relativistic gas was studied by Sorkin et al.~in \cite{Sorkin:1981wd}; additional work on self-gravitating gas solutions includes \cite{Tolman:1939jz,Oppenheimer:1939ne,Bondi:1947fta,Buchdahl:1959zz,1964PhRv..136..571M,Bowers:1974tgi,Herrera:1997plx,Andreasson:2007ck}. Papers with particular focus on the entropy of these gases include \cite{Gao:2011hh,Kim:2019ygw}; see Sec.~\ref{subsec:max_s}.

Ultra-compact objects that can optically mimic black holes are widely known, see \cite{Cardoso:2019rvt,Bambi:2025wjx} for a comprehensive review. As for thermodynamic mimics, it was pointed out in \cite{Banks:2002fj} that maximally stiff stars \cite{Zeldovich:1961sbr} have entropy of order $M^2$. The same is true of frozen stars \cite{Brustein:2018web,Brustein:2021lnr,Brustein:2023hic,Bambi:2025wjx} (see also \cite{Sorkin:1981wd}). It is argued in \cite{Brustein:2023hic} that frozen stars are perfect mimics, with exactly $S=4\pi M^2$, though this requires a transition region with transverse pressure much greater than its energy density.  Consistent with this, it is found in \cite{Riojas:2026god} that a class of systems, including the above, can precisely match the entropy of a black hole; but (excepting the HBH) the dominant energy condition is violated at their edges, due to their internal radial pressure.  We will say more about these examples in later sections.
Meanwhile the thermodynamics of black holes surrounded by thin shells of matter has been explored broadly, for instance in \cite{Martinez:1996ni,Andre:2019zzo,Andre:2021ctu,Lemos:2023yiz}; see also \cite{Riojas:2026god}.

In AdS space, a fast-rotating black hole can be surrounded by a disk of radiation at the same temperature; these have been called ``grey galaxies'' \cite{Kim:2023sig}. However, in this case the entropy of the ``galaxy'' is parametrically smaller than that of the black hole, of order $1/\sqrt{G}$. If there is a direct connection between this galaxy and the (non-rotating) solution considered here,  we have not yet found it.\footnote{While contemporary artificial intelligence tools were used for literature review, they are not responsible for the ideas and results in this paper.}

\section{Temperature of an HBH}\label{sec:thermo}

Having explored the classical properties of HBH metrics in Paper I \cite{Riojas:2026ytt}, we now consider their thermodynamic properties, under the strong assumption that the black hole and the ocean are thermally coupled and in a quasi-equilibrium state. (We will return to the limitations of this assumption, and others, in Sec.~\ref{sec:meanings}.) Strikingly, the HBH solution  mimics a black hole not only optically (as we saw in Paper I) but also in temperature.\footnote{In \cite{Brustein:2021lnr,Brustein:2023hic} it is argued that the frozen star solution of \cite{Brustein:2018web,Brustein:2021lnr},  which replaces a black hole of mass $m$ with an exotic material of total mass $m$, has the correct temperature to mimic a black hole of mass $m$.}

With $f(r)\neq j(r)$ at the horizon, and $\kappa$ the surface gravity, the Hawking temperature for the HBH metric in Eqs.~\eqref{fjmetric0}-\eqref{massfunction0} is
\begin{equation}\label{THLargeR}
    T_{H} = 
    \frac{\kappa}{2 \pi} = 
    \frac{1}{4\pi}\sqrt{\frac{df}{dr}\frac{dj}{dr}}\Bigg|_{r=2m}=
     \frac{1}{4\pi}\frac{m}{M}\frac{dj}{dr}\Bigg|_{r=2m}
    = ({8 \pi M})^{-1} \ .
\end{equation}
The surrounding ocean redshifts the temperature $(8 \pi m)^{-1}$ of the central black hole of mass $m$. Remarkably, the Hawking temperature is independent of $m$, and matches that of a black hole of mass $M$, due to a unique feature of this metric: $f(2m)/j(2m)=m^2/M^2$.  This same result can be obtained via a standard Bogoliubov  transformation; see \cite{Riojas:2026god}.

The local Tolman temperature is
\begin{equation}
    T(r) \equiv \frac{1}{\beta(r)}= \left(8 \pi M \sqrt{f(r)}\right)^{-1} = \left({8 \pi {\widehat{m}}(r)}\sqrt{1-\frac{2{\widehat{m}}(r)}{r}}\right)^{-1}
\end{equation}
for all $r$. 
Explicitly,  the local temperature near the black hole takes the form
\begin{equation}
    T(r<3m) = \left({8 \pi m}\sqrt{1-\frac{2m}{r}}\right)^{-1}
    \ ,
\end{equation}
but within the ocean it has the $m$- and $M$-independent power-law form 
\begin{equation}\label{Tshell}
    T(r) = \frac{\sqrt{27}}{8 \pi r} =\frac{\sqrt{3}}{8\pi \widehat{m}(r)}
 \ .
\end{equation}
This then glues on smoothly to the usual black-hole $T(r)$ formula at $r>3M$; see Fig.~\ref{fig:Metric}.  
Note that \eqref{Tshell} is the same temperature found in \cite{York:1986it} for a black hole in a cavity, when the cavity radius $r=3M$ coincides with the black hole's photon sphere.  This will be relevant in Sec.~\ref{subsec:York}.

The local temperature $T(r)$ must apply both to quanta in the ocean and to the Hawking radiation at the same radius.  In particular, the surfaces of the ocean have the same temperature as the Hawking modes there.\footnote{While the high-energy tail of the radiation from an HBH and black hole of mass $M$are the same, an attentive observer will notice that the gray-body factors differ and depend on $m$. The relevant effective potential is discussed in Paper I.}  

Mimicking the temperature of a black hole is a non-trivial feature of the NC gas that fails in generic systems.
For instance, shells of null matter that are formally in thermal equilibrium with a black hole \cite{Martinez:1996ni,Andre:2019zzo,Andre:2021ctu,Lemos:2023yiz} have non-minimal, $m$-dependent temperatures, unless they form a continuous null cluster. 

Consider a single shell of traceless radiation at radius $r_0$ surrounding a black hole of mass $m$, where $M\gg m$ is the ADM mass of the system.  In the limit $m\to 0$ this would be a Wheeler geon \cite{PhysRev.97.511,Misner:1973prb}, with $r_0\to \frac94 M$.  Outside the shell, the metric is Schwarzschild, while for $r<r_0$ it has the form
\begin{equation}
    f(r) = \frac{1-2M/r_0}{1-2m/r_0}(1-2m/r) \ \ , \ \ j(r) = 1-2m/r \ \ 
\end{equation}
The Hawking temperature is redshifted by the square root of the prefactor in $f(r)$, giving
\begin{equation}
    T_{H,shell} = T_H\sqrt{\frac{1-2M/r_0}{1-2m/r_0}}\approx \frac{1}{24\pi m} \gg \frac{1}{8\pi M}
\end{equation}
up to order $m/M$ corrections, with $M$ not appearing at leading order.

As we will see later, if the NC gas is replaced with a subluminal Einstein cluster, the temperature becomes $m$-dependent and warmer,  see Eq.~\eqref{Tdelta}. A cosmological constant also makes the temperature of the HBH different from that of a corresponding black hole, see Sec.~\ref{sec:AdS}.

If (as discussed in \cite{Riojas:2026god}) an HBH is approximated as a series of nested shells of matter, the continuous HBH is the only one whose temperature is $m$-independent and as low as that of a black hole of mass $M$. This is proven, in the special case of shells of traceless matter, in Appendix~\ref{app:coldestNC}. A more powerful and general result, covering a much broader set of shell configurations and employing heat capacities as a central tool, is given in \cite{Riojas:2026god}; one consequence is that among all spherically symmetric configurations of mechanically stable thin shells, the Hawking temperature at asymptotic infinity is minimized by the HBH.

\section{Power-law Thermodynamic Relations}\label{sec:powers}

We now consider whether the entropy of an HBH might also match that of a black hole.  A first question is whether it should be expected that the entropy of self-gravitating fluid might actually scale as $S\propto M^2$. The formal answer, well-established in the literature, is yes, even for a ball of gas with no central black hole. 

\subsection{Review of Anisotropic TOV equation}\label{subsec:derivation}

First, we briefly review the anisotropic TOV equation and a set of well-known self-similar solutions to that equation that are useful here.  For more details, see Paper I. The TOV equation generalizes simply to $d>4$, as do the HBH solutions of Paper I; see Appendix~\ref{app:otherHBH}. 

Under the static and spherically symmetric ansatz for the metric
\begin{equation}\label{metricform}
    ds^2 = -f(r) dt^2 + \frac{1}{j(r)} dr^2 + r^2 d\theta^2 + r^2 \sin^2\theta \ d\phi^2\ , 
\end{equation} 
along with the anisotropic stress-energy tensor, written in \textit{mixed} coordinates, of the form
\begin{equation}
    T_{\nu}^\mu = (-\rho, P_r, P_\perp, P_\perp) \quad \quad G^\mu_\nu+\Lambda \delta^\mu_\nu=8 \pi G T^\mu_\nu \ ,
    \label{eq:mixedeinsteinequation}
\end{equation}
the Einstein equations in the presence of a possible cosmological constant $\Lambda$ yield 
\begin{equation}
    \frac{f'(r)j(r)}{r  f(r)}-\frac{1-j(r)}{r^2 }=8 \pi  P_r -\Lambda
    \label{eq:EErrcomponent}
\end{equation}
\begin{equation}
  -\frac{j^{\prime}(r)}{ r}+\frac{1-j(r)}{r^2}=  \frac{1}{r^2}\frac{d}{d r}\left[r\left(1-j(r)\right)\right]=8 \pi \rho + \Lambda \ .
    \label{eq:EEttcomponent}
\end{equation}
Integrating the latter gives
\begin{equation}
    j(r) = 1-\frac{2 \widehat{m}(r)}{r}-\frac{\Lambda r^2}{3} \quad \quad \widehat{m}^{\prime}(r)=4 \pi r^2 \rho(r).
   \label{eq:jr}
\end{equation}
These imply
\begin{equation}
    \frac{r f'(r)}{2 f(r)}=\frac{{\widehat{m}}(r)+4 \pi  r^3 P_r-\frac{1}{3}\Lambda  r^3}{r-2 {\widehat{m}}(r)-\frac{1}{3}\Lambda  r^3}.
    \label{eq:EErrwitha(r)}
\end{equation}
The Bianchi identities then imply energy conservation:
\begin{equation}
    P_r^{\prime}(r)=-\frac{f^{\prime}(r)\left(P_r(r)+\rho(r)\right)}{2 f(r)}-\frac{2\left(P_r(r)-P_{\perp}(r)\right)}{r}\ .
    \label{eq:hydrostaticequilibrium}
\end{equation}
Combining these equations, we obtain the anisotropic TOV equation with a cosmological constant:
\begin{equation}
    P_r^{\prime}(r)=-\left(P_r(r)+\rho(r)\right) \frac{{\widehat{m}}(r)+r^3\left(4 \pi P_r(r)- \frac{1}{3}\Lambda\right)}{r(r-2 {\widehat{m}}(r)-\frac{1}{3}\Lambda r^3)}-\frac{2\left(P_r(r)-P_{\perp}(r)\right)}{r}.
    \label{eq:anisotropic_TOV}
\end{equation}

If $\Lambda=0$, self-similar solutions can be obtained by taking 
\begin{equation}\label{selfsimilardensities}
    P_r=w_r\rho\ , \ P_\perp=w_\perp \rho\ , \ m(r)=\frac{\nu r}{G}= 4\pi r^3 \rho(r)\ \  \Rightarrow \ \ \rho = \frac{\nu}{4\pi Gr^2}\ ,
\end{equation}
where $w_r$, $w_\perp$, $\nu$ are constants. Meanwhile, within the gas,
\begin{eqnarray}\label{generaldelta}
  j(r)=1-2\nu \ \ &,& \ \ f(r)\propto r^\delta \ \ , \ \ \nonumber \\[8pt] \nu=\frac{2w_\perp}{(1+w_r)^2+4w_\perp} \ , \  \delta &=& \frac{2\nu}{(1-2\nu)}(1+w_r) = \frac{4w_\perp}{1+w_r} \ ; 
\end{eqnarray} 
note $j(r)$ is constant. For ordinary gases with positive energy density and non-negative pressure, the non-negative parameter $\nu\leq\frac13$ is a function of $w_r,w_\perp$ (see Paper I \cite{Riojas:2026ytt}).  For a traceless gas, with $2w_\perp=1-w_r$,
\begin{equation}\label{bwnulllinearmass}
    \nu = \frac{1-w_r}{3+w_r^2} \ \ ; \ \ \delta={\frac{2(1-w_r)}{1+w_r}} \ \ .
\end{equation}
For an Einstein cluster ($w_r=0$), representing an almost non-interacting gas of objects  with speed $v^2/c^2=2w_\perp$ moving on randomly-oriented circular orbits, we have
\begin{equation}
    \nu=\frac{2w_\perp}{1+4w_\perp} \ ; \ \delta=4 w_\perp = \frac{2\nu}{1-2\nu} \ .
\end{equation}

 As for an NC ocean, we have $w_r=0$, $w_\perp =\frac12$, $\nu=\frac13$ and $\delta=2$; this solution also exists for non-zero $\Lambda$.  Within the ocean, the mass function, density and pressure are
\begin{equation}\label{gassmassfunction}
    {\widehat{m}}(r) = \frac{r}{3G} \ \  \Rightarrow \ \ \rho(r) = 2P_\perp(r) = \frac{1}{12\pi Gr^2} \ .
\end{equation}
This NC gas makes up the ocean of the HBH; see Fig.~\ref{fig:Metric} and Eqs.~\eqref{fjmetric0}-\eqref{massfunction0}.

If $\delta=2$, the metric of the fluid forms an {\it extended photon sphere} (as defined in Paper I \cite{Riojas:2026ytt}) on which massless particles can travel on circular geodesics.  Of these, the case $\nu=1/3$ is special: the extended photon sphere is {\it aligned} with the natural photon sphere of the system at $r=3M$, in which case the fluid can itself be made of circularly orbiting massless particles.  We will refer to other extended photon spheres, with $\nu\neq 1/3$, as {\it misaligned}.

Fluids with $P_r\neq0$ will expand or contract radially unless, at each surface, there is either a constraining wall or a transition region in which the pressure and density drop smoothly to zero. Edge effects are considered in detail in \cite{Riojas:2026god}. Here we often restrict our calculations to the more closely related case of Einstein clusters, where edge effects can be avoided and self-similar solutions require no corrections (see Paper I \cite{Riojas:2026ytt}).

\subsection{Thermodynamic Power Laws for General Fluids }\label{subsec:detailedpowerlaws}

Sorkin, Wald, and Zhang (SWZ) \cite{Sorkin:1981wd}, using a principle of entropy maximization (which we will return to in Sec.~\ref{subsec:max_s}), obtained solutions to the TOV equation for a self-gravitating $d=4$ null isotropic gas in a box of radius $r_0$.
Though the solutions are not in closed form, key properties can be obtained analytically, including the temperature and entropy, knowing only ${\widehat{m}}/r$ and $d{\widehat{m}}/dr$ at $r=r_0$. For a box of radius $r_0\sim M$, the temperature at the box wall scales like $T(r_0)\sim r_0^{-1/2}$, while the entropy $S\sim r_0^{3/2}\sim M^{3/2}$, far below that of a corresponding black hole.  

The scaling laws for $T(r_0)$ and $S$, applying to the global properties of the gas, are also found {\it locally} for the self-similar solution for the same gas. That solution has $\delta=1$ and thus $f(r)\propto r$, so the local temperature and entropy density scale as
\begin{equation}
T(r)\sim \frac{T(r_0)}{\sqrt{f(r)}}\sim \frac{1}{\sqrt r} \ \Rightarrow
s(r) \sim T(r)/\rho(r)\sim r^{-3/2} \ .
\end{equation}
Integrating $s(r)$ gives $S\sim r_0^{3/2}\sim M^{3/2}$.

The standard undergraduate Boltzmann argument for a photon gas, $\rho+P = \left( d\rho/dT \right)_V$, can be applied to any isotropic substance. For a traceless gas, $P = \rho/d$, giving $\rho \sim T^{d}$, $s\sim T^{d-1}$,
while a more general equation of state $P=w\rho$, where $0<w\leq 1$, implies
\begin{equation}\label{isogasscaling2}
    \rho \sim T^{1+1/w} \ \ ; \ \ s\sim T^{1/w} \ \Rightarrow \ s\sim \rho^{1/(1+w)}  \ .
\end{equation}
(Note dimensional analysis assures  $\rho \sim Ts$ and $M\sim TS$.)
For $d=4$, a ball of radius $r_0$ with $M\sim \rho r_0^3$ and $S\sim s r_0^3$ then has 
\begin{equation}\label{isoSMr}
    S\sim M^{1/(1+w)} r_0^{3w/(1+w))} \ \ \ \ \ \ (d=4) .
\end{equation} 
If the ball is self-gravitating and relativistic so that $r_0\sim M$, then
\begin{equation}\label{isoSM}
    S\sim M^{(1+3w)/(1+w)} \ \ \ \ \ \ \ (d=4).
\end{equation}
This gives the expected answer for the traceless gas with $P=\rho/3$, and also shows that if $P=\rho$, the stiffest equation of state, we have $\rho\sim T^2$, $s\sim T$ \cite{Zeldovich:1961sbr} and $S\sim M^{2}$ \cite{Banks:2002fj}.

For the anisotropic case, the scaling laws for self-similar solutions in $d=4$ can be found using \eqref{selfsimilardensities} and \eqref{generaldelta}, along with $\rho\sim Ts$ by dimensional analysis:
\begin{equation}\label{Tnaivescaling}
    T(r)\sim f(r)^{-1/2}\sim r^{-\delta/2}
    \end{equation}
\begin{equation}\label{Snaivescaling} s\sim r^{-2+\delta/2} \Rightarrow S\sim r_0^{1+\delta/2}
\end{equation} 
The relation \eqref{Snaivescaling} was found by other methods in \cite{Kim:2019ygw}, with $\delta=1+2w_2/(1+w_1)$ in their notation; for additional details, see Sec.~\ref{subsec:max_s}.

Using the fact that $r_0\propto M$ for a self-similar solution, we then learn
\begin{equation}\label{SvsM}
  S  \sim M^{1+\delta/2} \ .
\end{equation}
Since $\rho\sim 1/r^2$ for any  self-similar solution,
\begin{equation}\label{generalgasscaling}
    \rho \sim T^{4/\delta} \  , \ \ s \sim T^{4/\delta-1} \ .
\end{equation} 
For $d=4$ isotropic gases, $4/\delta= 1+1/w$, in agreement with \eqref{isogasscaling2}.

Thus the relations found for isotropic gases  apply for all anisotropic self-similar fluids with the same $\delta$. (We will refine the relations between $s,\rho,T$ in Sec.~\ref{subsec:max_s}, at which point $\nu$-dependence appears.)
In particular, all materials with $ 2w_\perp=(1+w_r)$ have $\delta=2$ and $S\sim M^2$, including the NC gas with $w_r=0$, $\nu=\frac13$, the stiffest star \cite{Zeldovich:1961sbr, Banks:2002fj} with $w_r=w_\perp=1$, $\nu=\frac14$, and the frozen star with $w_\perp=0,w_r\to -1$ and 
$\nu\to\frac12$ \cite{Brustein:2018web,Brustein:2021lnr,Brustein:2023hic}.\footnote{See also \cite{Sorkin:1981wd}, who proposed $\nu=\frac12$ and concluded it has $S\propto M^2$, though they did not recognize the required anisotropy or obtain it as a solution to the anisotropic TOV equation.}  
In fact, in every $d\geq 4$, the stiff star, frozen star and NC gas always satisfy $\rho \sim T^2$, $s\sim T$, and have $S$ proportional to the area of the Schwarzschild-Tangherlini black hole in 
$d$ dimensions.\footnote{The power laws  $\rho \sim T^2$, $s\sim T$ for the NC gas are perhaps surprising; since the excitations making up the gas travel on $d-2$-spheres of definite radius, one might have expected the effective dimension of the gas to be reduced by one. Instead this suggests a hidden system with $d=2$.} 
See  Appendix~\ref{app:scaling} for certain $d>4$ scaling laws.

\begin{figure}
    \centering
    \includegraphics[width=0.75\linewidth]{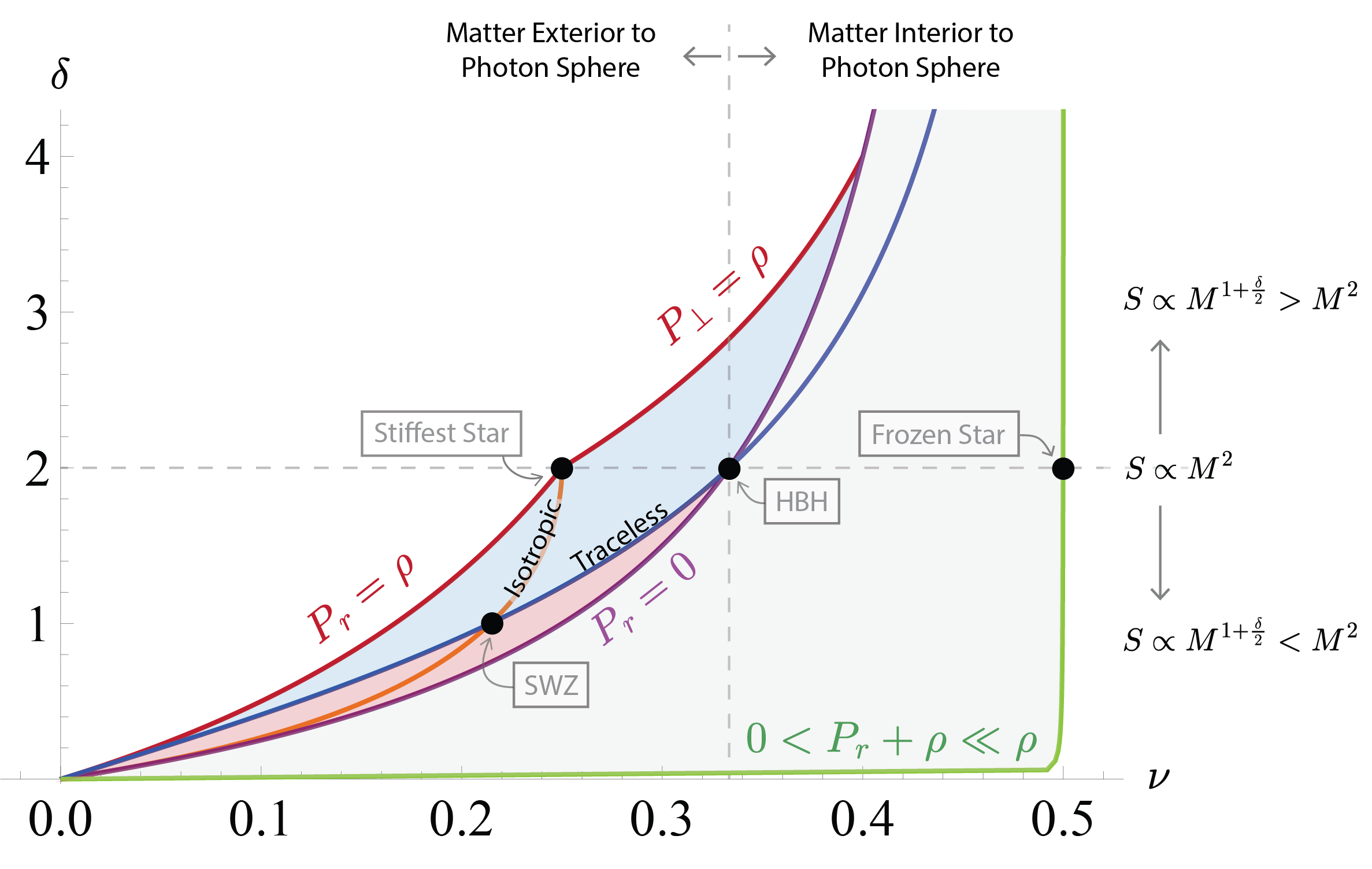}
    \caption{ Classes of self-similar matter with various $\nu$ and $\delta$ values. The smallest region (pink, center) is that of ordinary gases with $0\leq P_r,P_\perp,P_r+2P_\perp\leq\rho$; the upper shaded region (blue) is that of other substances with $0\leq P_r,P_\perp\leq\rho$, while the lower right (grey) is that of substances with $P_r<0$. Also shown are curves of isotropic gases, traceless gases, gases with $P_r=0$, and gases with $P_r\approx -\rho$. The HBH has $\nu=\frac13$; other substances with $\delta=2$ include the stiffest stars \cite{Zeldovich:1961sbr,Banks:2002fj} $(P=\rho)$  and frozen stars \cite{Brustein:2018web,Brustein:2021lnr,Brustein:2023hic} $P_r\approx -\rho, P_\perp=0$. Also shown is the isotropic traceless gas $(P=\rho/3)$ of \cite{Sorkin:1981wd} (SWZ). }
    \label{fig:nudelta}
\end{figure}

The space of self-similar solutions is sketched in Fig.~\ref{fig:nudelta}; see also Paper I. The curves for isotropic fluids, for Einstein clusters ($P_r=0$) and for traceless gases are shown, as are the two key dividing lines at $\nu=\frac13$ and $\delta=2$.  Also indicated are the null isotropic gas studied by Sorkin, Wald and Zhang (SWZ) \cite{Sorkin:1981wd}, as well as the fluids for the stiffest star, HBH and frozen star.

As noted above, self-similar fluids with $\delta<2$ have $S<M^2$. Those with $\delta=2$ have $S\sim M^2$ and create extended photon spheres.  Solutions with $\delta>2$ are likely  problematic, but we have not shown this in general. An Einstein cluster with $\delta>2$ involves superluminal motion, while a traceless gas with $\delta>2$ has negative radial pressure.  
Meanwhile, materials with $\nu<\frac13$ are not ultra-compact and have no photon sphere of their own; when placed around a black hole, they lie outside the black hole's photon sphere.  Those with $\nu>\frac13$ are ultra-compact, have negative radial pressure, and lie inside their own photon sphere.  The NC gas, at the intersection of the four curves $w_r=0$, $w_r+2w_\perp=1$,  $\nu=\frac13$ and $\delta=2$, has zero radial pressure and a traceless equation of state, creates an extended photon sphere that is aligned with the exterior Schwarzschild metric's photon sphere, and has $S\propto$ Area. These statements generalize to higher dimensions.

\subsection{Implications of Dimensional Analysis}

Dimensional analysis for the self-similar solutions with $\rho = \nu/(4\pi G r^2)$ and outer radius $r=r_0$, under the assumption that no scales enter except fundamental constants of nature, gives more refined scaling laws. Those for general $d$ and $\delta$ are given in Appendix~\ref{app:scaling}.

For an isotropic traceless gas,
\begin{equation}
     T\propto \frac{\hbar^{(d-1)/d}}{G^{1/d}\ r^{2/d}}\ , \ 
     \rho\propto \frac{T^{d}}{\hbar^{d-1}} \ , \ s\propto \frac{1}{(\hbar G)^{(d-1)/d}\ r^{2-2/d}} \Rightarrow S\propto 
     \left[\frac{A_{d-2}[M]}{G\hbar}\right]^{\frac{d-1}{d}}
\end{equation}
where $A_{d-2}[M]$ is the area of a black hole of mass $M$ in $d$ dimensions.
The relation between $\rho$ and $T$ is independent of $G$, consistent with the generalized Stefan-Boltzmann law, while fractional powers of $G$ appear when quantities are expressed in terms of $r$. The total entropy $S\sim M/T$ is parametrically less than that of a corresponding black hole.

By contrast, for the NC gas as well as other $\delta=2$ systems, we have 
\begin{equation}\label{NCNDA}
T\propto \frac{\hbar}{r} \ \ ; \ \ \rho\propto \frac{T^2}{\hbar^2G}\ \ ; \ \  s\propto \frac{1}{\hbar G r}  \ 
\Rightarrow \  S\propto
     \left[\frac{A_{d-2}[M]}{G\hbar}\right] \propto \frac {G^{\frac{1}{d-3}}M^{\frac{d-2}{d-3}}}{\hbar} \ ;
\end{equation}
thus the entropy $S$ scales as that of  a black hole in $d$ dimensions. Here  $\rho$ and $s$ are proportional to $1/G$, while the temperature $T$ is independent of $G$, as is true for a black hole's Hawking temperature near the photon sphere.

 The scaling $s\sim (\hbar Gr)^{-1}\sim \ell_{pl}^{-2}r^{-1}$ for the NC gas, while enormous, is 
 necessary. A self-similar material has $\rho\sim 1/Gr^2$, from \eqref{selfsimilardensities}, while equilibrium forces the temperatures of the NC gas  and Hawking radiation to match, implying  $T\sim \hbar/r$ from \eqref{Tshell}. The result follows from $s\sim \rho/T$.

All these expressions apply to self-similar balls with their  mild singularities at $r=0$.
If we place a $\delta\neq 2$ gas around a black hole with mass $m\ll M$, equilibrium  will require the temperature to depend on $m$ (see for instance Sec.~\ref{subsec:matteractionzero}), altering the dimensional analysis.   For an HBH, however, the temperature \eqref{THLargeR} is independent of $m$, and so the local scaling laws still apply, as do the global ones up to  order-($m/M$) corrections. Naively \eqref{THLargeR} applies to any $\delta=2$ system, but it may be altered by edge effects unless $P_r=0$.

\section{The relation between $\rho$ and $Ts$}\label{sec:rhoTs}

For an HBH of mass parameters $\{m,M\}$, it is easily checked that if $\rho=Ts$, and if the entropies of the black hole and gas add, then the total entropy matches that of a black hole of mass $M$.  For instance, in $d=4$ using \eqref{gassmassfunction} and \eqref{THLargeR}, this gives
\begin{equation}\label{rhoTimpliesS}
    s(r)=\frac{\rho(r)}{T(r)} = \frac{2}{9\sqrt{3}Gr} \ \ \Rightarrow 
\ \
    S_{ocean} = \int_{3Gm}^{\hat{3GM}} dr\ 4\pi r^2 \sqrt{g_{rr}}\  s(r) = 4\pi G(M^2-m^2) \ ,
\end{equation}
exactly what must be added\footnote{Note $S_{ocean}$ is {\it smaller} than one-fourth the oriented area of the ocean's outer and inner surfaces by a factor of $(2/3)^2$: the squared ratio of the radii of the horizon and photon sphere.} to the entropy $4Gm^2$ of a black hole of mass $m$ to match the entropy $4GM^2$ of a black hole of mass $M$.  Analogous results hold in all $d\geq 4$.

Checking the coefficient of the HBH entropy thus requires checking that $Ts/\rho = 1$ exactly.  For an isotropic gas in $d$ dimensions, the result is instead $Ts/\rho = d/(d-1) = (1+w_r).$  We will now argue, using an entropic variational principle, that the general relation
\begin{equation}\label{rhoprTs}
    (1+w_r)\rho = \rho+P_r=Ts
\end{equation}
holds for all $d$, $w_r$, $w_\perp$. We will see an independent argument, using the Euclidean path integral, in Sec.~\ref{subsec:York}.

\subsection{Local Euler Relation from Entropy Maximization}\label{subsec:max_s}

In SWZ \cite{Sorkin:1981wd}, solutions to the TOV equation for an isotropic gas were derived by requiring the entropy be maximized using a variational principle. The principle has also been extended  for more general isotropic fluids \cite{Gao:2011hh} and considerable progress has been made on anisotropic fluids \cite{Kim:2019ygw}. The anisotropic scaling relations \eqref{Snaivescaling}, and more generally \eqref{eq:kimandlee} below,  were obtained by Kim and Lee \cite{Kim:2019ygw}, who determined $s(\rho(r),r)$.

Here we apply the method to infer a wider set of thermodynamic relationships that hold for a more general anisotropic but spherically-symmetric gas.  Our methods, which are somewhat parallel to those of \cite{Kim:2019ygw}, directly use the local temperature.

Taking  a static, spherically symmetric energy momentum tensor of the form \eqref{fjmetric0}, let us assume we have a solution to the anisotropic TOV equation. As reviewed in Sec.~\ref{subsec:derivation}, the equation is derived from Eqs.~\eqref{eq:EErrcomponent} and \eqref{eq:EEttcomponent}, the ($rr$) and ($tt$) components of the Einstein equation, along with the Bianchi identity which requires  local stress-energy conservation.
The solution consists of a mass function $\widehat m(r)$, a density $\rho(r) =4\pi r^2 \widehat m'(r)$, and radial and transverse pressures $P_r(r), P_\perp(r)$, along with a metric with functions $j(r)$, $f(r)$.  From the latter  we can determine the radial dependence of the temperature up to a normalization constant, since $T(r)\propto 1/\sqrt{f(r)}$. 

The entropy maximization principle is that the total entropy $S$ should be a local maximum with respect to functional variations of $\widehat m(r)$ and $\widehat m'(r)=4\pi r^2\rho(r).$  In other words, the entropy density, written as $s[\rho(r),r]$, should satisfy an Euler-Lagrange equation, following from the maximization of the integral
\begin{equation}
    I_S=\int s[\rho(r),r]\ \sqrt{h} \ d^3 x = 4 \pi \int\frac{s[\rho(r),r]}{\sqrt{j[\widehat m(r),r]}} r^2 d r.
\end{equation}
where $h$ is the spatial part of the metric. (The entropy density $\hat s$ should depend only on $\rho\propto \widehat m'$, not $\widehat m$ itself, since a change in the mass of a central object far from the gas should not change the gas thermodynamics.)
In the isotropic gas case of SWZ, one takes $s[\rho,r]=\rho^{3/4}$ since $\rho\sim T^4, s\sim T^3$; there is no explicit $r$ dependence.  We will see that explicit $r$ dependence is only required in anisotropic situations.

Varying ${\widehat{m}}(r)$, and imposing the boundary condition at $r=r_0$ that $\widehat{m}(r_0)=M$, we get Euler-Lagrange equations:
\begin{equation}
    \delta_{\widehat{m}} I_S=\int\left[\left(\frac{\partial s}{\partial \rho}\right)\left(\frac{r^2}{\sqrt{j}}\right) \frac{\delta\left(d{\widehat{m}}/dr\right)}{4 \pi r^2}+s \frac{\partial\left(\frac{r^2}{\sqrt{j}}\right)}{\partial {\widehat{m}}} \delta {\widehat{m}}\right] 4 \pi d r.
\end{equation}
We next integrate by parts and make the usual assumption that $\delta {\widehat{m}}=0$ at the edges of the interval in order to drop the boundary term. We take $j(r)$ to be defined as in \eqref{eq:jr}: it contains the term $-2{\widehat{m}}/r$ with no other ${\widehat{m}}$ dependence, so $\partial j/\partial{\widehat{m}} = -2/r$. Finally, assuming that $(\partial s(r)/\partial\rho(r))|_r = 1/T(r)$, we find
\begin{equation}
    -\frac{d}{d r}\left(\frac{1}{T(r)\sqrt{j(r)}} \right)=4 \pi r\left(j[r]\right)^{-3 / 2} s(r).
\end{equation}
This can be rewritten as
\begin{equation}
    \frac{T'(r)}{T(r)}=\frac{-\frac12  \frac{\partial j(r)}{\partial r} + 4\pi rs(r)T(r)}{j(r)} \ , 
\end{equation}
or, using $T(r)\propto 1/\sqrt{f(r)}$,
\begin{equation}\label{sTvsfj}
     sT = \frac{1}{8\pi}\frac{j}{r}\left[\frac{f'}{f}-\frac{j'}{j}\right] \ .
\end{equation}
Notice this is always zero in vacuum (for static spherically symmetric solutions) since there $f\propto j$.

The right hand side of Eq.~\eqref{sTvsfj} is the sum of the $(rr)$ and $(tt)$ components of the Einstein tensor, Eqs.~\eqref{eq:EErrcomponent}-\eqref{eq:EEttcomponent}, multiplied by $1/8\pi$.  We therefore obtain
\begin{equation}\label{localthermo}
     sT = T_r^r - T_\mu^\mu = P_r+\rho \ .
\end{equation}
This confirms that $P_r=0$ implies $\rho=Ts$, and thus verifies \eqref{rhoTimpliesS}:  the entropy of the NC cluster is exactly the difference between the entropy of a black hole of mass $M$ and one of mass $m$. 

The general form of $j(r)$ for spherically symmetric matter with a cosmological constant $\Lambda$ is $j=1-2{\widehat{m}}/r -\frac13 \Lambda r^2$.  In this case
\begin{equation}
    \frac{\partial j(r)}{\partial r} = \frac{2}{r^2}\left(-r{\widehat{m}}'+ {\widehat{m}}-\frac13\Lambda r^3\right) \ .
\end{equation}
This gives us
\begin{equation}
    {\frac{T^{\prime}}{T}=\frac{r {\widehat{m}}'-{\widehat{m}}+\frac13 \Lambda r^3 -4 \pi r^3(s T)}{r(r-2 {\widehat{m}}-\frac13 \Lambda r^3)}}
\end{equation}
We then can reproduce Eq.~\eqref{eq:EErrwitha(r)},
\begin{equation}
    \frac{ f'}{2 f}=\frac{{\widehat{m}}+4 \pi  r^3 P_r-\frac13 \Lambda r^3}{r(r-2 {\widehat{m}}-\frac13 \Lambda r^3)} \ ,
    \label{eq:dlogfEq}
\end{equation}
if we use $T'/T=-f'/2f$ and if
\begin{equation}
(4\pi r^3)sT=r{\widehat{m}}'+4\pi r^3P_r = 4\pi r^3(\rho+P_r) \ .
\end{equation}
Thus $sT=\rho+P_r$ is true independent of $\Lambda$. 

Note that for the interesting case of $\delta=2$ where the total entropy $S\sim M^2$, this equation reads $sT=\rho(1-2\nu)/\nu$. This implies that as $\nu\to 1/2$, either $s\to 0$ or $T\to 0$; thus there is no local entropy density if the local temperature is finite.  (The region $\nu\to 0$ for $\delta=2$ has infinite radial pressure and is unphysical, see Fig.~\ref{fig:nudelta}.)

Taking the radial derivative of $s T = P_r + \rho$, we find
\begin{equation}
    \frac{d P_r}{dr} + \left[ \frac{d \rho}{dr} - T \frac{ds}{dr}\right] = s \frac{dT}{dr}.
\end{equation}
This resembles the condition for energy conservation: 
\begin{equation}
    \frac{dP_r}{dr} + \left[\frac{2(P_\perp - P_r)}{r}\right] =-\frac{f^{\prime}(r)\left(P_r(r)+\rho(r)\right)}{2 f(r)}=s \frac{d T}{d r} .
\end{equation}
Combining these two equations and using
\begin{equation}
    \frac{ds}{dr}=\left(\frac{\partial s}{\partial \rho}\right)_r \frac{d \rho}{dr} + \left( \frac{\partial s}{\partial r}\right)_\rho = \frac{1}{T} \frac{d \rho}{dr} + \left( \frac{\partial s}{\partial r}\right)_\rho
\end{equation}
gives the following thermodynamic equation:
\begin{equation}
    \frac{d \rho}{d r}-T \frac{d s}{d r} = -T\left( \frac{\partial s}{\partial r}\right)_\rho= \frac{2\left(P_{\perp}-P_r\right)}{r}\ .
    \label{eq:masterformulascaling}
\end{equation}
Note this implies that $s$ depends explicitly on $r$ only in anisotropic gases.  Also, since $(\partial \rho/\partial s)_r=1/T$, we obtain a differential thermodynamic relation for an anisotropic fluid: 
\begin{equation}
    \boxed{
    d \rho = T ds + (P_\perp - P_r) d \log A, \quad \quad A \equiv 4 \pi r^2.
    }
\end{equation}
These relations hold for any spherically symmetric $P_r(r)$ and $P_\perp(r)$. 

In the special case $P_r=w_r \rho$, $P_\perp = w_\perp \rho$ where $w_r,w_\perp$ are constants, we can use
\begin{equation}
    \frac{\partial \log s}{\partial \log \rho} = \frac{1}{T}\frac{\rho}{s}=\frac{1}{1+w_r} \ \ , \ \ 
    \frac{\partial \log s}{\partial \log r} = -\frac{2\left(P_{\perp}-P_r\right)}{Tr}\frac{r}{s} = -2\frac{(w_\perp-w_r)}{1+w_r}
\end{equation}
where we substituted \eqref{localthermo} in the first expression and \eqref{eq:masterformulascaling} in the second.
From this we recover the relations found by Kim and Lee \cite{Kim:2019ygw}:
\begin{equation}\label{eq:kimandlee}
    s\propto \rho^{1/(1+w_r)} r^{-2(w_\perp-w_r)/(1+w_r)} .
\end{equation}
For self-similar solutions with $\rho\propto 1/r^2$, this becomes \eqref{generalgasscaling}:
\begin{equation}
    s \propto \rho^{(1+w_r-w_\perp)/(1+w_r)} = \rho^{1-\delta/4}\ .
\end{equation}

\subsection{Global Euler Relations}

If we take a ball of gas, and locally we have the Euler relation
$\rho + P_r=Ts$, this should have implications for the global thermodynamics of the gas in terms of its mass $M$, its total entropy $S$, and the temperature $T_\infty$ as measured at infinity. Here we obtain a global Euler relation by integrating the local Euler relation derived in the previous section.
A self-similar solution in $d=4$ has $r_0=M/\nu$ and $\rho=\nu/(4\pi r^2)$. (Here we take $G=\hbar=1$). 
If we define constants $c_{S,T}$ via $s=c_S/(4\pi r^{2-\delta/2})$, $T=c_T/r^{\delta/2}$, then $c_Tc_S=\nu(1+w_r)$.  We have 
\begin{equation}
    T(r_0) = \frac{c_T}{r_0^{\delta/2}} = T_\infty \frac{1}{\sqrt{g_{tt}(r_0)}}
\end{equation} 
as the surface temperature, and a total entropy 
\begin{equation}    
S=\int_{0}^{{M/\nu}} dr\ 4\pi r^2 \sqrt{g_{rr}}\  s(r) 
= c_S \sqrt{g_{rr}(r_0)} \frac{r_0^{\delta/2+1}}{\frac{\delta}{2}+1}
\end{equation}
where we have used the fact that $g_{rr}(r_0)=1/g_{tt}(r_0)=1/(1-2\nu)$ and $g_{rr}$ is constant within the self-similar solution. Using $g_{rr}g_{tt}=1$ for $r\geq r_0$, this gives 
\begin{equation}
    T_\infty S= \frac{M(1+w_r)}{1+\frac{\delta}{2}} \ .
\end{equation}
Thus
\begin{equation}\label{MTS}
    M = \frac{1+\frac{\delta}{2}}{(1+w_r)}T_\infty S  = \frac{(1+w_r+2w_\perp)}{(1+w_r)^2}T_\infty S 
\end{equation}
For an isotropic self-gravitating gas with $w_r=w_\perp=\frac13$, SWZ \cite{Sorkin:1981wd} found $M_{gas}=\frac{8}{9}T_\infty S_{gas}$, in agreement with our formula.

For the NC gas, \eqref{MTS} gives $M=2T_\infty S$, just as for a black hole.  Indeed black holes and  NC gases always match; the same argument for general $d$ gives   $M/TS = (d-2)/(d-3)$.

The case with walls is worked out in \cite{Riojas:2026god}. There it is shown\footnote{The full expression for the Euler relation in any static spherically symmetric spacetime is \cite{Riojas:2026god}: 
\begin{equation}
    S=\frac{A_{\mathrm{H}}}{4}+\sum_{\text {shells }} \frac{1}{2 T_{\infty}}\left[r \sqrt{f j}\left(\frac{r f^{\prime}}{2 f}-1\right)\right]_{R_i}+\frac{1}{2 T_{\infty}} \int_{\text {gas }} r \sqrt{f j}\left(\frac{f^{\prime}}{f}-\frac{j^{\prime}}{j}\right) d r .
\end{equation}
This formula, which has massless walls as a special case, computes coarse-grained entropies without the Euclidean path integral. The first term is the usual horizon area and the last term is related to Eq.~\eqref{sTvsfj}.} that for self-similar fluids with a {\it massless} exterior wall:
\begin{equation}
    2 T_{\infty} S=\left(\frac{4+w_r(2-\delta)}{\delta+2}\right)M,
\end{equation}
which for $\delta=2$ reduces to $M=2T_\infty S $, mimicking a black hole; but of course such walls badly violate the dominant energy condition. Only the NC gas, where such walls are unnecessary, exhibits mimicry compatible with the dominant energy condition.

\section{Vanishing of the Matter Action}\label{sec:nomatteraction}

In the previous section we found that $\rho(r)=T(r)s(r)$ for a $P_r=0$ gas.  Obtaining $s(r)$ from the known temperature and density, and  assuming thermal equilibrium across the HBH, we saw in \eqref{rhoTimpliesS} that 
\begin{equation}\label{SBHplusSocean}
    S_{BH}(m) + S_{ocean}(m,M) = 4\pi G M^2 = S_{BH}(M)
\end{equation}
for all $m$. 
However, as yet we have not shown 
\begin{equation}\label{EntropiesAdd}
S_{HBH}(m,M) =    S_{BH}(m) + S_{ocean}(m,M)
\end{equation}
which would establish our central result: the identity of HBH and BH (coarse-grained) entropies at the same mass.

In the introduction we sketched an argument that $S_{HBH}(m,M)=S_{BH}(M)$, based on the claim that their temperatures and Euclidean actions are equal.  The equality of the temperatures was shown in Sec.~\ref{sec:thermo}.  We will see the actions are equal for the following reasons.  First, the bulk gravity action $\int \sqrt{g}R$ vanishes because the NC gas is traceless, which ensures the Ricci scalar is zero.  The Euclidean matter action of the NC gas is also zero, as shown in Sec.~\ref{subsec:matteractionzero}. Finally, the HBH requires no walls at the ocean's surfaces, as its extended photon sphere is aligned.  For all three contributions to separately vanish requires the key properties of the NC gas: it is traceless and has $P_r=0$.  The only term that remains is the usual GHY boundary term, which is the same for a Schwarzschild black hole of mass $M$ and an HBH of the same mass.
With this result in mind, we will then directly compute the Euclidean action in Sec.~\ref{subsec:York} using a method of York \cite{York:1986it}, where the HBH is placed in a cavity. Not only does the result confirm \eqref{SBHplusSocean} and \eqref{EntropiesAdd}, it also shows that the entropy density $s(r)$ is non-zero only inside the ocean and agrees with \eqref{rhoTimpliesS}.  We can therefore be confident in the simple picture that the entropy is located only at the horizon and in the NC gas, and that those entropies add. 

In general systems of black holes coupled to matter shells, Martinez and York argued that their entropies should add, and furthermore that the matter action on the shell should be proportional to the free energy density \cite{Martinez:1989hn}.  In our case, the constituents of the NC gas are massless particles with no chemical potential, so the free energy density would be expected to be ${\bf f} = \rho - Ts$, which from \eqref{rhoprTs} is equal to $-P_r$.  We will show that, for the HBH, the matter action vanishes if and only if $\rho=Ts$ and the coarse-grained entropies add.  Specifically, in Sec.~\ref{subsec:matteractionzero} we show that $\rho=Ts$ and the assumption that entropies add implies a vanishing matter action in any self-similar Einstein cluster.  Meanwhile we show in Sec.~\ref{subsec:York} that a vanishing matter action for the NC gas solution implies that $\rho=Ts$ in the gas and that the entropies add for the HBH solution.

\subsection{Direct Check of the Entropy and Action}\label{subsec:matteractionzero}

We first check that $\rho=Ts$ implies a vanishing matter action {\it without} having to assume that the matter action is proportional to ${\bf f}$, or that ${\bf f}=\rho-Ts$, although we do assume \eqref{EntropiesAdd}.  

Consider a black hole of mass $m$ surrounded by a self-similar Einstein cluster with parameter $\nu$, whose metric is given in Appendix~\ref{app:otherHBH}. The matter extends from $r=Gm/\nu$ to $r=GM/\nu$, where $M$ is the ADM mass.  The temperature is set by 
\begin{equation}\label{Tdelta}
    T_{H} = \frac{1}{4\pi}\sqrt{\frac{df}{dr}\frac{dj}{dr}}\Bigg|_{r=2Gm}=
     \frac{1}{4\pi}\left(\frac{m}{M}\right)^{\delta/2}\frac{dj}{dr}\Bigg|_{r=2Gm}
= \frac{1}{8 \pi GM^{\delta/2}m^{1-\delta/2}} \ .
\end{equation}
Since  $m\leq M$ and $\delta\leq 2$ (to avoid $v>c$), the HBH is again unique: the coldest of all self-similar Einstein clusters surrounding a black hole.

Within the gas, $f(r) = (\nu r/GM)^\delta (1-2\nu)$, and the entropy density is
\begin{equation}
    s(r) = \frac{\rho(r)}{T(r)} = \frac{\rho(r)\sqrt{f(r)}}{T_H} ={2G^{-\delta/2}}{} \frac{\nu^{1+\delta/2}m^{1-\delta/2}}{ r^{2-\delta/2}}\sqrt{ 1-2\nu}
\end{equation}
Notice this depends on $m$ but not $M$.  The total entropy of the gas is then
\begin{eqnarray}
    S_{ocean} &=& \int_{Gm/\nu}^{GM/\nu} dr \frac{4\pi r^2}{\sqrt{j(r)}}s(r)
    =\frac{8\pi G}{1+\delta/2}\left({M^{1+\delta/2}m^{1-\delta/2}}-m^2\right) \ .
\end{eqnarray}
With a black hole of entropy $4\pi Gm^2$, and assuming the entropies add, we then have
\begin{equation}
    S_{total} = S_{ocean}+S_{BH} = 
     {4\pi G} \left({\frac{2}{1+\delta/2}M^{1+\delta/2}m^{1-\delta/2}}-\frac{1-\delta/2}{1+\delta/2}m^2\right) \ .
\end{equation}
For $M=m$, when there is no gas, we recover $S_{total} = 4\pi Gm^2/\hbar$, while for $\delta=2$, we recover what we claim for an HBH: $S_{total} = 4\pi GM^2/\hbar$. 
For $\delta<2$, the entropy and entropy density go to zero as $m\to 0$; this is because $T_H, T(r)\to \infty$ (with $\rho(r)$ fixed) in that limit.

We can compute the action from $F=I/\beta = E-TS$, where $E=M$ for the system as a whole and $T=T_H$. This gives
\begin{eqnarray}\label{IGHYbulk}
    I_{total} &=& \beta M-S_{total} = {8 \pi G}M^{1+\delta/2}m^{1-\delta/2}-{4\pi G} \left({\frac{2}{1+\delta/2}M^{1+\delta/2}m^{1-\delta/2}}-\frac{1-\delta/2}{1+\delta/2}m^2\right)
    \nonumber \\ &=&
    {4 \pi G}M^{1+\delta/2}m^{1-\delta/2}-{4\pi G} \frac{1-\delta/2}{1+\delta/2} \left({M^{1+\delta/2}m^{1-\delta/2}}-m^2\right)\nonumber \\ &=&
 I_{GHY} + (I_{ocean+bulk}) 
\end{eqnarray}
where $I_{GHY}$ is the usual Gibbons-Hawking/York boundary term for an asymptotically Schwarzschild metric with this value of $T_H$.  This gives the expected answers for $M=m$ as well as for $\delta=0,2$. However, at this stage we cannot separate the gas and bulk contributions to the second term.

For these self-similar Einstein clusters, the trace of the stress tensor is
\begin{equation}
T_{\mu}^{\mu} = \rho-2P_\perp\propto 1-2w_\perp = 1-\delta/2
\end{equation}
The bulk gravity action is proportional to the Ricci scalar, which is itself proportional to $T_{\mu}^{\mu}$.    For the Euclidean action, we have 
\begin{equation}
I_{bulk} =   - \frac{1}{16\pi G}\int\sqrt{-g}R
\end{equation}
For $P_r=0$, we have $f(r)\propto (r/r_0)^\delta j(r)$, where $r_0=GM/\nu$, implying
\begin{equation}
    R = \frac{(2-\delta ) \delta }{2 (1+\delta ) r^2} \ \ \ \ \ \ \ \ (P_r=0) 
\end{equation}
within the gas. Thus
\begin{equation}
    I_{bulk} = \frac{\beta}{4G}\frac{(2-\delta ) \delta }{2 (1+\delta ) }\int_{Gm/\nu}^{GM/\nu} dr \left(\frac{r\nu}{G M}\right)^{\delta/2} = 
    {4 \pi G}\frac{1-\delta/2 }{1+\delta/2}
    \left(m^{1-\delta/2} M^{1+\delta/2} -m^2\right)
\end{equation}
Comparing with \eqref{IGHYbulk} we see. 
\begin{equation}
    I_{ocean}=I_{ocean+bulk}-I_{bulk}=0
\end{equation}
implying that the matter action vanishes for all $P_r=0$ gases, including the NC gas. As we pointed out earlier, this is consistent with the suggestion of Martinez and York \cite{Martinez:1989hn}.

\subsection{Application of York's Method}\label{subsec:York}

Now we run the logic in another order: we compute HBH thermodynamics assuming that the GHY boundary term gives the entire contribution to the Euclidean action.  In doing so, we assume the matter action is zero, though not that the matter action is proportional to $\rho-Ts$ nor that $\rho-Ts=0$.  Instead it {\it follows}  that $\rho=Ts$ uniformly within the ocean, that $dS/dr\propto s$ is non-zero only within the ocean, and that both \eqref{SBHplusSocean} and  \eqref{EntropiesAdd} are true.

To do this, we implement the approach of York \cite{York:1986it} in which we place the HBH in a cavity of radius $r_0$ whose temperature is fixed at $T(r_0)$, and reanalyze its thermodynamics with Dirichlet boundary conditions.\footnote{Dirichlet boundary conditions are suitable for our problem but they are not elliptic \cite{2006math.....12647A,An:2021fcq,Witten:2018lgb}. For different purposes, alternative boundary conditions can be used. Conformal boundary conditions are elliptic; a relevant Hamiltonian, $E^{\text{CBC}}$, was given in \cite{Odak:2021axr}. It was later specialized and interpreted at finite cutoff in the conformal canonical ensemble in Appendix A of \cite{Banihashemi:2024yye}, where it was noted that $E^{\text{CBC}}$ changes sign when the black hole branches merge. It is interesting to note that this energy vanishes precisely when the cutoff surface sits at the photon sphere, where $r f'(r_{\textrm{ps}})/2f(r_{\text{ps}})=1$, as well as throughout the ocean of the HBH. This geometric fact holds independently of the thermodynamic interpretation in \cite{Banihashemi:2024yye}. We leave an exploration of the HBH, and the photon sphere, in the conformal canonical ensemble to future work.} 
As before, only the boundary GHY term contributes to the action, now computed at the finite radius $r_0$. The usual background subtraction approach regulates the calculation by subtracting off thermal flat space with the appropriately selected temperature \cite{Gibbons:1976ue,York:1986it}. We relegate the details to Paper III, where we will perform the more general calculation in AdS$_4$. The results of that calculation are reproduced in Appendix~\ref{app:AdSYorkAction}; taking the asymptotically-flat limit, one obtains the formulas of this section.

The result for the action boils down to this: simply take the results found by York in \cite{York:1986it} and replace everywhere the Schwarzschild mass $M$ with the  
mass function ${\widehat{m}}(r_0)$: the mass of the HBH that is contained within York's boundary, given in  equation \eqref{massfunction0}. 
Specifically, with a boundary at $r_0$ and a boundary inverse temperature $\beta(r_0)$, York's famous black hole action
\begin{eqnarray}\label{BHaction}
    I_{BH}(M,r_0) &=& 12 \pi M^2 - r_0 [8 \pi M -\beta(r_0)]
\end{eqnarray}
becomes, for the HBH, simply
\begin{eqnarray}\label{HBHaction}
    I_{HBH}(m,M;r_0) &=& 12 \pi [{\widehat{m}}(r_0)]^2 - r_0[8 \pi {\widehat{m}}(r_0) -\beta(r_0)]
\end{eqnarray}
The action at large (small) $r_0$ is that of the corresponding black hole of mass $M$  $(m)$.  The action in the ocean interpolates between the two as follows: 
\begin{equation}
    I_{HBH}(3m<r_0<3M) = -\frac{4\pi r_0^2}{3}\left(1-\frac{2}{\sqrt{3}}\right) 
\end{equation}
A scale-invariant result, independent of both $m$ and $M$, is to be expected, since the GHY calculation only knows the metric at $r=r_0$, where the metric is a scale-invariant cone.

The system's energy is given by
\begin{eqnarray}
E(r_0)= \frac{\partial I}{\partial\beta(r_0)}\Bigg|_{\widehat{m}} + \frac{\partial I(r_0)/\partial {\widehat{m}}(r_0)}{\partial \beta(r_0)/\partial {\widehat{m}}(r_0)}\Bigg|_{\beta}
&=&
r_0 \left( 1 - \sqrt{1-\frac{2{\widehat{m}}(r_0)}{r_0}}\right)
\end{eqnarray}
This is what York finds for a Schwarzschild black hole with ${\widehat{m}}$ replaced with $M$. 
We thus match \cite{York:1986it}  at large $r_0$ (and find the corresponding result for $r_0<3m$), while in the ocean we obtain
\begin{equation}
    E(3m<r_0<3M) = r_0\left(1-\frac{1}{\sqrt{3}}\right)  ,
\end{equation}
again a scale-invariant interpolation between the small- and large-$r$ regions.
Also, as in \cite{York:1986it}, 
\begin{equation}\label{mvsE}
{\widehat{m}}(r_0) = E(r_0) -
\frac{1}{2r_0} E(r_0 )^2
\ \ 
\Rightarrow
\ \ 
dE(r_0) = \frac{1}{ \sqrt{1-\frac{2{\widehat{m}}(r_0)}{r_0}}} \  d{\widehat{m}}(r_0) \ \ \ \ [{\rm fixed\ }r_0] .
\end{equation}
due to self-energy corrections.  

The entropy contained within the sphere of radius $r_0$ is then
\begin{equation}
    S(r_0) = \beta(r_0) E(r_0) - I(r_0) = 4\pi [{\widehat{m}}(r_0)]^2 \ .
\end{equation}
Thus $S=4\pi m^2$ for $r_0<3m$, $S=4\pi M^2$ for $r_0>3M$, and with $r_0$ sitting within the ocean,
\begin{equation}
    S(r_0) = \frac{4}{9}{\pi r_0^2}
    = 4\pi m^2 + 4\pi \left( \frac{r_0^2}{9}-m^2\right)
\end{equation}
steadily growing with $r_0$ until $r_0=3M$ and $S(r_0)=4\pi M^2$. This is a check of Eqs.~\eqref{SBHplusSocean} and \eqref{EntropiesAdd}.

From this we can obtain the entropy density.  Outside the ocean the entropy is piecewise constant, so $s=0$ there.  Within the ocean, 
\begin{equation}\label{sdefn}
\frac{    \partial S(r_0)}{    \partial r_0}\Bigg|_{\widehat{m}} = (4\pi r_0^2 \sqrt{g_{rr}}) s(r_0)  \ \ \Rightarrow \ \ s(r_0)  =\frac{2}{9\sqrt{3}r_0} \ , 
\end{equation}
confirming \eqref{rhoTimpliesS}, which was obtained via different logic.\footnote{For an HBH in equilibrium, the entropy density, and thus the total entropy of the ocean $S_{ocean}(m,M)$,  can be obtained by combining equation (47) of Kim and Lee \cite{Kim:2019ygw} with two additional pieces of information obtained here: the HBH temperature \eqref{THLargeR}, and the relation \eqref{localthermo} which implies $\rho=Ts$ for an NC cluster.}

 Meanwhile,
from  \eqref{mvsE} one obtains 
\begin{equation}
\frac{    \partial S(r_0)}{    \partial {\widehat{m}}(r_0)}\Bigg|_{r_0} = \frac{1}{T(r_0)}
\frac{    \partial E(r_0)}{    \partial {\widehat{m}}(r_0)}\Bigg|_{r_0}
\end{equation}
a check of the first law of HBH thermodynamics. Then
\begin{equation}
    T\frac{dS}{dr_0} = T\frac{dS}{d{\widehat{m}}}\frac{d{\widehat{m}}}{dr_0} = \frac{1}{8\pi M \sqrt{f(r_0)}}(8\pi {\widehat{m}}) {\widehat{m}}' = \frac{{\widehat{m}}'}{\sqrt{j(r_0)}} \ .
\end{equation}
which implies
\begin{eqnarray}\label{rhoTs}
    \rho(r_0) = \frac{1}{4\pi r_0^2} {\widehat{m}}'(r_0) =
    \frac{1}{4\pi r^2} \sqrt{j(r_0)} T(r_0) \frac{d S}{dr_0}   = T(r_0) s(r_0) \ .
\end{eqnarray}
Thus we find that assuming the matter action vanishes implies ${\bf f} = \rho-Ts=-P_r = 0$.

Let us now turn to the heat capacity $C_A$ of these systems.  For an ordinary black hole in a spherical cavity  \cite{York:1986it},
\begin{equation}\label{specificheat}
    C_A=\frac{\partial E}{\partial T}\Bigg|_{A_0} = -T\frac{\partial S}{\partial T}\Bigg|_{A_0} = 8\pi M^2\frac{r_0-2M}{r_0-3M}\ ,
\end{equation}
diverging to $-\infty$ as $r_0\to 3M$ from above and to $+\infty$ as $r_0\to 3M$ from below.  The divergence occurs because the temperature, viewed as $T(r_0,E)$, is minimized when $E$ is appropriate to a black hole of mass $M=r_0/3$. Thus the temperature is independent, to first order, of the energy for fixed $r_0=3M$, implying $\partial \beta(r_0)/\partial E   = 0$ at the photon sphere. 
Thus a black hole is unstable thermodynamically in any box with $r_0>3M$.

For the HBHs, we find the same result with $M$ replaced with ${\widehat{m}}(r_0)$.  Now  the specific heat is positive and divergent as $r\to 3m$ from below, and negative and divergent as $r\to 3M$ from above. Within the ocean, it is ill-defined: better said, $1/C_A$ is zero.  We can trace this fact back to 
\begin{equation}\label{dbetadmhat}
\frac{\partial \beta(r)}{\partial {\widehat{m}}(r)}\Bigg|_r = 
\frac{8 \pi  [r-3 {\widehat{m}}(r)]}{r \sqrt{1-\frac{2 {\widehat{m}}(r)}{r}}} \ .
\end{equation}
Thus the temperature $T(r_0,E)$ is at its minimum when the ocean extends to $r=r_0$, implying $\widehat{m}(r_0)=r_0/3$.  This makes a large ocean act like a heat bath: an infusion or extraction of energy does not change the ocean's temperature profile $T(r)$, and instead makes it grow larger or smaller.

\subsection{Interpreting the Original York calculation}\label{subsec:YorkInterpret}

It has been noticed \cite{Andre:2021ctu} that there is a relation between York's original calculation for a black hole and the calculation of a shell around a black hole carried out by Frauendiener et al.~\cite{Frauendiener:1990nao} and by Brady et al.~\cite{Brady:1991np}.  Here we provide an interpretation and use it to find further evidence for the vanishing of the matter action for an HBH.  Additional  interpretation is provided in \cite{Riojas:2026god}.

 York notes that a complete thermodynamic description of the system requires not only the variables $E,T,S,A_0=4\pi r_0^2$ but also a quantity conjugate to $A_0$, which he calls the surface pressure.  (He uses ``$\sigma$'' as notation for this quantity, but we will refer to it as $p_Y$, for ``York pressure.'')  York's first law of thermodynamics, and corresponding Euler relation, are: 
\begin{equation}\label{York1st}
    dE = T\ dS - p_Y\ dA_0, \quad \quad  E = 2 (TS -p_Y A_0).
\end{equation}
One might at first glance expect that $p_Y$, conjugate to area, would be related to $P_r$ or $P_\perp$ in a fluid solution, but we will see in a moment that this is not the case. 

York calculates $p_Y$ for a black hole, and for an HBH we find the same result with $M\to {\widehat{m}}(r_0)$:
\begin{equation}
    p_Y = \frac{1}{8\pi r_0}\left(\frac{1-\dfrac{\widehat{m}}{r_0}}{\sqrt{1-\dfrac{2{\widehat{m}}}{r_0}}}-1\right)
\end{equation}
In the ocean this takes the form
\begin{equation}
    p_Y = \frac{1}{8\pi r_0}\left(\frac{2}{\sqrt{3}}-1\right)
\end{equation}

\begin{figure}
    \centering
    \includegraphics[width=0.7\linewidth]{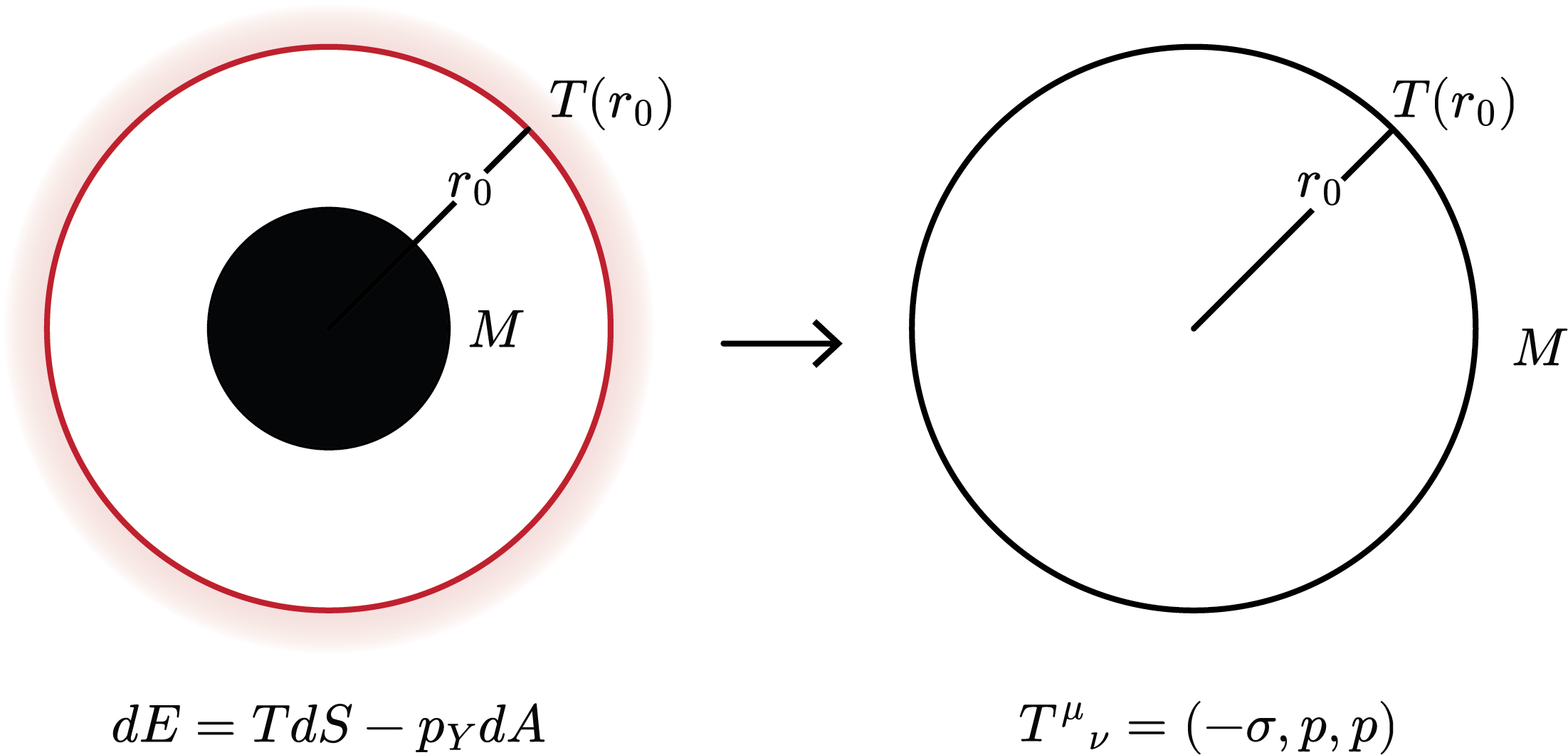}
    \caption{At left, York's cavity of radius $r_0$ and temperature $T$ surrounding a black hole of mass $M$; the system has energy $E$ and surface pressure $p_Y$. To this we associate the system at right: a shell of matter of ADM mass $M$ at temperature $T$ with energy surface density $\sigma$ and transverse pressure $p$.}
    \label{fig:YorkCavityToShell}
\end{figure}

Let us return to a black hole of mass $M$ and replace it, as in Fig.~\ref{fig:YorkCavityToShell}, with an infinitesimal shell at $r=r_0$ with the same mass and temperature. This shell will be static if it satisfies the Israel junction condition \cite{Israel:1966rt} with zero mass interior to the shell \cite{Frauendiener:1990nao,Brady:1991np}. These conditions on the shell's energy density $\sigma$ and two-dimensional isotropic pressure $p$ are directly related to the quantities $E$ and $p_Y$ in York \cite{York:1986it}:
\begin{equation}\label{sigmapEApy}
    \sigma =\frac{1}{4\pi r_0} \left(1-\sqrt{1-\frac{2M}{r_0}}\right) = \frac{E}{A_0} , \quad \quad p = \frac{1}{8\pi r_0}\left(\frac{1-\frac{M}{r_0}}{\sqrt{1-\frac{2M}{r_0}}}-1\right)=  p_Y
\end{equation}
Thus York's calculation identifies the values of $p,\sigma$ needed for a thin shell to be static at the cavity wall at temperature equal to the local Hawking temperature $T_{shell}=T(r_0) =T_H/\sqrt{1-2M/r_0} $.

 If this is true, then the free energy of the shell at $r=r_0$ should equal York's free energy \eqref{BHaction}.  By this we mean the free energy obtained from the action of the shell's matter, with any GHY term only measuring additional matter interior to $r=r_0$. The shell has no radial pressure, by construction, so we expect no free-energy-like contribution on the shell itself; as we showed in Sec.~\ref{subsec:matteractionzero}, there is no gas action for Einstein clusters, presumably because ${\bf f}=-P_r=\rho-Ts=0$.  However, we expect a non-zero trace contribution from the shell to take the place of the $R_\mu^\mu$ contribution in Sec.~\ref{subsec:matteractionzero}. Writing the energy-momentum tensor localized to the shell as $^{(3)}{T}_\mu^\nu = {\rm diag}(\sigma, p,p)$, the shell contribution to the Euclidean action is 
\begin{eqnarray}
    \frac{1}{2} \int_0^{\beta_H} \sqrt{g_{tt}}\ dt\int\  ^{(3)}{T}_\mu^\mu\  r_0^2\  d\Omega\  = 
    \frac{\sigma-2p}{2 G}(4\pi r_0^2)\beta(r_0)
\end{eqnarray}
which indeed, using \eqref{sigmapEApy}, reproduces equation \eqref{BHaction}. See \cite{Riojas:2026god} for more extensive application of this and similar results.

A specific case of interest is $r_0=\frac94 M$, the Buchdahl radius.   At this radius, a shell made of traceless radiation should be static; such a shell is a Wheeler geon \cite{PhysRev.97.511,Misner:1973prb,Andreasson:2021lsh,Andreasson:2015agw}.  As noted but not elaborated on by York, the free energy of the black hole in the cavity vanishes for this value of $r_0$. Here we see that the tracelessness of the radiation on a shell at that radius implies that there is no classical contribution to the action.  Together with our observation that ${\bf f}=-P_r=0$ for any shell, we conclude that York's result at the Buchdahl radius supports our claims about NC gas.  

It has been noted \cite{Andre:2021ctu} that this feature of the York action at the Buchdahl radius persists in all dimensions.  Indeed, a Wheeler geon for any $d\geq 4$ should always have zero matter action, just as for anything made of NC gas.

\subsection{Construction by Nested Shells}

A complementary derivation of HBH thermodynamics is given in \cite{Riojas:2026god}, where the system is constructed by building it up layer-by-layer. For thin shells in asymptotically flat spacetime, the regions of thermodynamic and mechanical stability are highly constrained; they overlap only for infinitesimally light shells, located at the photon sphere \cite{York:1986it,Brady:1991np}, with traceless stress-energy tensors. The HBH is then realized dynamically as the unique solution to the joint stability problem for thin layers of self-gravitating matter. In the limit of vanishing radial pressure, the anisotropic TOV equation organizes self-gravitating matter into nested Israel layers. The $(q,\mu)$ fixed point in SWZ \cite{Sorkin:1981jc} becomes a line of fixed points, and analytic solutions to the TOV equation cross this line at points that recover the Israel junction conditions. Continuous solution curves are tangent to the critical line; for traceless matter the single continuous case is the HBH ocean, located at the photon sphere with $\widehat{m}(r)=r/3$.

Black hole thermodynamics are then derived directly from the Israel junction conditions, by building up the geometry layer-by-layer while tracking the asymptotic temperature. The finite-radius specific heat is closely related to the thermodynamic rules for building the system: adding an infinitesimal shell at fixed total mass cools or heats the configuration according to the sign of $C_A^{-1}$, while $C_A^{-1}=0$ leaves the asymptotic Hawking temperature unchanged. Adding a shell at fixed radius is interpreted as adding heat to the system, deriving York's local thermodynamics without using the Euclidean path integral or entropy maximization. Finally, it is shown that at the photon sphere, $C_A^{-1} \propto \Lambda$ in all dimensions. Applied shell by shell to the HBH, this gives the same entropy relation obtained above $S_{HBH}(m, M)=S_{BH}(M)$, providing an explanation for why our results match those of York, but with $M$ replaced with $\widehat{m}$. It also recovers the usual results for ordinary black holes by constructing them layer-by-layer from self-gravitating matter.

\section{Black Holes and HBHs in AdS$_4$ space}\label{sec:AdS}

We continue now with a brief discussion of AdS hillingar black holes, in which we briefly summarize some of the main results of Paper III \cite{RiojasStrasslerAdS}.  In the canonical ensemble, the equilibrium-HBH model shows only subleading changes to the standard story, but the microcanonical ensemble is substantially altered.\footnote{We are deeply grateful to D.~Jafferis for extensive discussions concerning certain subtle issues.}

In Paper I \cite{Riojas:2026ytt}, we exhibited HBH solutions in AdS space, reproduced in Appendix~\ref{app:otherHBH}.  These similarly have a black hole of mass $m$ and horizon radius $r_+$ surrounded by NC gas, such that the whole system has mass $M$.  Let us specialize to AdS$_4$, with cosmological constant $\Lambda=-3/L^2$ and  (where  $1-2m/r_+ +r_+^2/L^2=0$); the ocean extends from $r=3m$ to $r=3M$ as in flat space. Results in higher dimensions are analogous; those in de Sitter follow from the exchange $L^2\to-L^2$.  

A simple calculation shows that the AdS$_4$-HBH temperature is {\it not} independent of $m$, in contrast to flat space:
\begin{equation}\label{AdSHBHTemp}
    T_H =  \frac{\lambda_M}{\lambda_m} T_m 
    \ \ ; \ \ T_m =   \frac{ \left(L^2m +r_+^3\right)}{2\pi L^2  r^2_+}
    \equiv \frac{1}{\beta_m}, 
\end{equation}
where $T_m$ is the Hawking temperature of an AdS$_4$ black hole of mass $m$ and 
\begin{equation}\label{Lyapunovs}
    \lambda_M^2 = \frac{1}{27M^2} + \frac{1}{L^2}  \ ;
      \quad \quad  \lambda_m^2 = \frac{1}{27m^2} + \frac{1}{L^2} 
    \ .
\end{equation}
The local temperature $T_H/\sqrt{f(r)}$ goes to zero at the AdS boundary, so  we use the usual rescaled temperature  \cite{Witten:1998qj}
\begin{equation}\label{AdSHBHTemplocal}
    T_{}(r)=T_H\frac{\sqrt{f_{AdS}(r)}}{\sqrt{f_{HBH}(r)}},
\end{equation}
where the extra factor of $f_{AdS}$ ensures $T(r\to\infty) = T_H$.  In a York-style calculation in which the HBH is placed in a cavity, the same rescaling also ensures that the energy in the cavity approaches the ADM mass $M$, rather than zero, as $r_0\to\infty$. 

One can then compute the coarse-grained entropy using the three methods we have used in flat space: the Euclidean action and entropy maximization explained above, and the method of nested shells explained in \cite{Riojas:2026god}.  All methods yield
\begin{equation}\label{AdSHBHEntropy}
    S(m,M,L)  = S_{BH}(m,L) + S_{ocean}(m,M,L) =
  \pi r_+^2 + \left[\lambda_m\beta_m L^2 (M\lambda_{M}-m\lambda_m) \right].
\end{equation}
Using York's method as we did in Sec.~\ref{subsec:York}, we find as above that the entropy density is only non-zero in the ocean and that the division above into $S_{BH}$ and $S_{ocean}$ is consistent.  The Euclidean action for an AdS$_4$-HBH in a cavity is given in Appendix~\ref{app:AdSYorkAction}.

Computing the free energy from  $F =  M - T_HS$, using \eqref{AdSHBHTemp} and \eqref{AdSHBHEntropy}, or by direct calculation of the Euclidean action, we find the main results of \cite{Hawking:1982dh} are unchanged.  Specifically, at low temperatures the dominant state is that of a thermal gas in AdS space, while at high temperatures above the Hawking-Page temperature, a large black hole dominates.  

However, subdominant saddle points undergo important changes.   First, relative to small black holes with $m=M_S\ll L$, the continuum of HBHs with $m<M<M_S$ at the same temperature have free energy that decreases with $m$, with the minimum as $m\to 0$.  Second, at temperatures just above the spinodal temperature $T_{spin}=\sqrt{3}/(2\pi L)$, where the dominant state is the thermal gas in AdS space, there are two black holes of almost equal free energy.  In this regime, the NC gas has free energy {\it lower} than that of both black holes at the same temperature.
We will briefly explore some implications of these facts in Sec.~\ref{subsec:cavitycanon}.

Moreover, there are HBH states even below the spinodal temperature, as can be seen  from \eqref{AdSHBHTemp}; these are HBH states with  $m\ll L\ll M$.  The state with minimum temperature $T_{min}=\sqrt{27}/(8\pi L)$ has $m\to 0, M\to \infty$, in which NC gas fills the entire space.

The effect on the microcanonical ensemble is more dramatic: a black hole of mass $M$ has lower entropy than pure NC gas of the same mass.\footnote{The reverse is true in de Sitter space.}  
This implies that {\it if} thermodynamic equilibrium is possible, AdS black holes are thermodynamically unstable in the microcanonical ensemble: an AdS black hole of mass $M$ can potentially cool by emitting radiation, eventually becoming a colder ball of NC gas of mass $M$.  This suggests that in this system, the canonical and microcanonical ensembles are incompatible (if equilibrium is possible in the first place).

Specifically, from Eq.~\eqref{AdSHBHEntropy},
\begin{equation}\label{AdSHBHEntropy2}
    S_{BH}(M)\equiv S(M,M,L)  = \pi R_+^2 \ \ ; \ \
    S_{gas}\equiv S(0,M,L) = \frac{8\pi}{{27}}L^2 (\sqrt{27}M\lambda_{M}-1) 
\end{equation}
where  $1-2M/R_+ +R_+^2/L^2=0$.
These are equal in flat space $(L\to\infty)$, in agreement with our earlier results, but for finite $L$ the black hole always has {\it smaller} entropy.  The reason that NC gas wins over black hole horizons in AdS, but not in flat space, seems to be that AdS black holes have smaller-area horizons than those in flat space, since $r_+<2m$,  while  their photon spheres have the same area.

A small black hole and a corresponding HBH both have $T_H\sim 1/M$, $S_{BH}\sim M^2$, and $M\approx 2T_H S$ as in flat space; the difference in their entropies is parametrically small:
\begin{equation}\label{SHBHvsSBH}
S_{HBH}(M,m) - S_{BH}(M) = \left(\frac{ 27M^4 + 10m^2 M^2 -5 m^4}{L^2}\right) + {\cal O}\left(\frac{M^6}{L^4}\right) > 0\ .
\end{equation}
For a large black hole, $T_H\sim M^{1/3}$ and $S_{BH}\sim M^{2/3}$, with $M\approx \frac{2}{3}T_HS$.  However, for a pure NC gas with $M\gg L$, we have $T_H\to T_{min}\sim 1/L$, a constant as $M\to \infty$, and $S_{gas}\sim ML\gg S_{BH}$: the entropy difference is parametrically large.

In fact, the equations above imply $M=T_{min}S$, with coefficient unity, as $M,S\to\infty$.  This is characteristic of Hagedorn-like scaling, with $T_{min}$ serving as a critical temperature, which implies an ill-defined canonical ensemble.  Such  unexpected behavior may only indicate that we have pushed the toy model beyond its region of physical applicability and that equilibrium is not possible in this regime.  It could also indicate that
Hagedorn-like scaling is modified by corrections, which is a common occurrence in systems with first-order phase transitions with large latent heat.  For instance, quantum corrections might  reduce the power of $M$ slightly (by order-$1/N_c^2$ in any CFT$_{d-1}$ dual description), allowing the canonical ensemble to be well-defined.  

However, such a small change would still leave  large black holes thermodynamically unstable to evaporate to NC gas.  If one believes that large black holes ought to be microcanonically stable --- in the language of the large-$N_c$ CFT on a sphere, that the hot deconfined phase of the theory should not spontaneously cool and reorganize into an unfamiliar, high-entropy state --- then one might well conclude the toy model is simply unphysical. However, there are other possibilities, as we will explore further in Paper III. 

A possibly important subtlety, potentially relevant for any AdS/CFT applications of the equilibrium-HBH system, is that certain thermodynamic questions about the system are sensitive to a cutoff at large $r$ ({\it i.e.}, to an ultraviolet cutoff in any dual quantum field theory). Suppose we place the system in a York-style cavity of radius $r_0\gg L$. Then among states of fixed energy $E$, the state with largest entropy is the NC gas that fills the cavity; this has $E\sim M=r_0/3$ and $S\sim r_0 L$, which far exceeds the entropy of an AdS black hole of the same mass. Thus the microcanonical ensemble is dominated by gas states.  However, among systems that can fit within the cavity --- that is, states that have radius $r\leq r_0$ --- the state with the largest entropy is a black hole that completely fills the cavity, $S\sim \pi r_0^2$, which far exceeds $r_0L$. This would allow the canonical ensemble to be dominated by black holes.  We will consider  these delicate issues further in Paper III.\footnote{A similar point arises even in flat space, where in a cavity of radius $r_0=3M$, a black hole of mass $M$ and a cavity-filling NC gas  of mass $M$ have equal entropy and dominate the microcanonical ensemble, while a black hole of mass $3M/2$ fills the cavity and dominates the canonical ensemble.}

\section{A Black Hole in a Small Box}\label{sec:cavity}

Now we turn to the following question: within the equilibrium-HBH toy model, what is the evolution of a black hole in a small spherical cavity?  We will explore this both in the microcanonical and canonical ensembles.  In the former case, we find that a black hole can evaporate inside a box whose radius is smaller than the usual minimum,  $r_0^{min} \sim M^{5/3}$.  Instead the minimum radius in flat space is at the photon sphere, $r_0^{min}=3M$. In the latter case, we revisit some expectations concerning the evolution of certain subleading saddle points.

Before we begin, let us recall one more observation of York \cite{York:1986it} concerning a black hole in a cavity of radius $r_0$ with  a fixed temperature $T_0$ at the cavity wall and zero cosmological constant.  For sufficiently large $T_0$, York showed that the black hole can have two possible masses $M_S$ and $M_L$, with $M_S < r_0/3 < M_L$. From \eqref{specificheat}, the large black hole is thermodynamically stable in the canonical ensemble, but the small black hole is not. 
Only one black hole exists at the temperature $T=\sqrt{27}/(8 \pi  r_0)$, with none at lower temperatures. (It is not accidental, as we will explore in Paper III \cite{RiojasStrasslerAdS}, that this is reminiscent of the situation in AdS spaces without a York boundary.)  This ``spinodal'' temperature at the transition is precisely the local temperature of the ocean seen in \eqref{Tshell}, and the spinodal black hole's photon sphere sits at the cavity wall.  

 Since, within York's cavity, a small black hole's photon sphere has $r=3M_S<r_0$, a traceless ocean may be established extending from $3M_S$ to any $r$ not exceeding $r_0$.  Thus the small black hole of mass $M_S$ connects to a continuum of HBH solutions with $m=M_S<M\leq r_0/3$, all of which have negative specific heat. The large black hole's photon sphere lies beyond York's boundary, and so cannot host a traceless ocean; it is a stable but isolated solution with no attached HBH continuum.

\subsection{Microcanonical Ensemble}\label{subsec:cavitymicro}

We begin with the microcanonical ensemble for a black hole in a cavity, fixing $M$ and maximizing the total coarse-grained entropy $S$ subject to that constraint.  (More precisely, we fix the total energy $E$ in the box, which has the effect of fixing the HBH metric parameter $M$ when we are in flat space.) The temperature at the cavity wall is not fixed, and we take the cavity wall to be a perfect reflector so as to avoid any dissipation. 

We first recall standard lore concerning the microcanonical ensemble for a black hole in a cavity whose Hawking radiation becomes ordinary null isotropic gas.  Assume first that $r_0\gg M$, so that the black hole volume is small; then (recall Sec.~\ref{subsec:detailedpowerlaws}) $d=4$ isotropic gas filling the cavity has entropy
 $S_{gas}=\zeta \ (m_{gas}\ r_0)^{3/4}$, where $\zeta$ is an order-one constant.  The black hole can evaporate entirely to the isotropic gas only if the final gas entropy $S_{gas}^{(f)}$ exceeds the initial black hole entropy $S_{BH}^{(i)}$, i.e.
\begin{equation}\label{BH2isogas}
S_{BH}^{(i)}= 4\pi GM^2 < S_{gas}^{(f)}= \zeta \ (Mr_0)^{3/4}  \ \ \Rightarrow \ \  M\lesssim \left(\frac{\zeta}{\pi G}\right)^{4/5}\left(\frac{r_0}{2}\right)^{3/5} \ .
\end{equation}
Thus the black hole can evaporate in a cavity of radius $r_0\gtrsim M^{5/3}$ (or an infinite space), but in a  smaller cavity it can only evaporate partially before the entropy can no longer grow. For example, if the cavity has radius $r_0 \sim  M^{3/2}$, then evaporation stops once an amount $\sim M^{1/2}$ of isotropic gas has been radiated.\footnote{For $m_{gas}\ll M$, $S\approx 4 \pi (M-m_{gas})^2 + \zeta (m_{gas} r_0)^{3/4}$, while $dS/dm_{gas}\approx -8\pi M + \frac{3}{4} \zeta{r_0^{3/4}}{m_{gas}^{-1/4}}$. Setting $r_0\sim M^{3/2}$ and solving $dS/dm_{gas}=0$ yields the result.}

Such is the argument in flat space; in AdS$_4$ space the argument is similar \cite{Banks:1998dd,Horowitz:1999uv}. Roughly speaking, only black holes with $M\ll {\rm min}(r_0^{5/3},L^{5/3})$ can evaporate.   This differs from the stability transition in the canonical ensemble, for which black holes develop negative specific heat if $M<r_0/3$ in flat space and $M<M_{spin}\equiv L/\sqrt{27}$ in infinite AdS space. Similar logic applies for any $d\geq4$.

 But if HBH states can exist, then instead of surrounding a black hole with entropy-poor isotropic gas, we can instead use entropy-rich NC gas, which can potentially match or outmatch the black hole.
As we saw in \eqref{SHBHvsSBH}, in AdS$_4$ space the black hole can always evaporate to NC  gas, while in flat space a higher-order calculation would be needed to  determine whether the black hole or NC gas is more stable.

\begin{figure}
    \centering
    \includegraphics[width=0.75\linewidth]{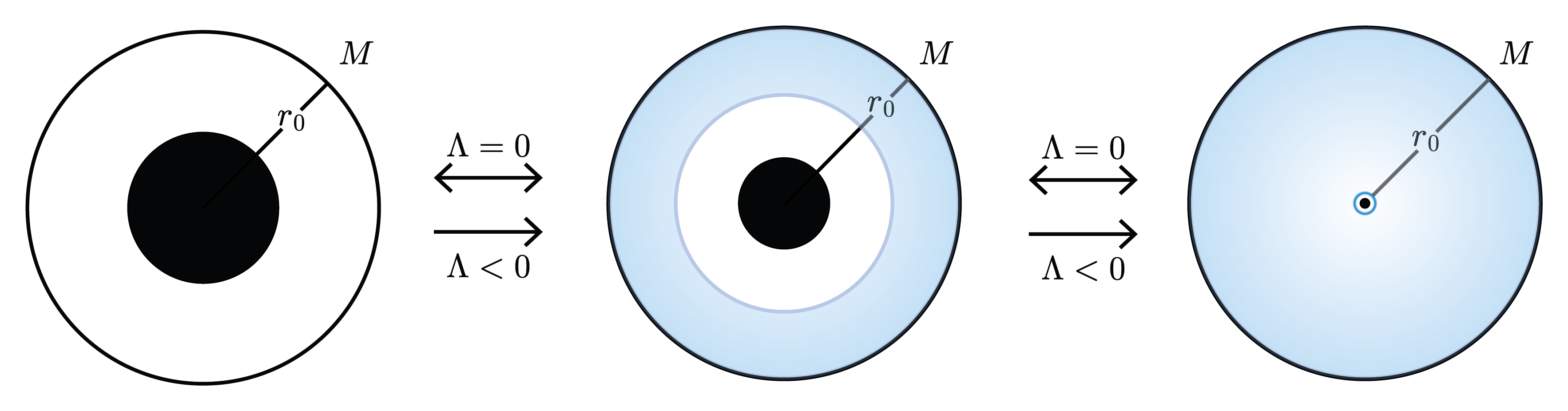}
    \caption{Black hole evolution within the microcanonical ensemble. In flat space, entropy for an HBH in a cavity of radius $3M$ is independent of $m$, while in AdS space it grows as $m$ decreases. An AdS black hole in such a cavity will thus become a ball of NC gas, while in flat space it is entropically allowed, but not required, to do so.}
    \label{fig:microcanonical}
\end{figure}

Particularly interesting is the case $M=r_0/3$, the spinodal black hole in flat space, which has zero inverse-heat-capacity and is thus marginally stable thermodynamically.  In this case, illustrated in Fig.~\ref{fig:microcanonical}, all HBH states with mass $M$ extend to the cavity wall,  with no vacuum region above the ocean; for $m\to 0$ the cavity is entirely filled with NC gas.  Since all of these states  have the same entropy independent of $m$, none is entropically favored over the others.

The degeneracy with $m$ raises a subtle point, which we now briefly address.  If $m$ were a zero mode of a pure state, like a $d=5$ instanton's  radius (though $m$ is compact) or its $SU(2)$ orientation (though $m$ is not a symmetry parameter), we would need to quantize $m$ as a collective coordinate. One might think that the ground state of the system involves a superposition over $m$, and that a mixed state that initially has $m\lesssim M$ will oscillate, with $m$ descending toward zero and returning; thus the black hole in a cavity may undecay after decaying, as long as coherence is not lost. But $m$ may be an emergent zero mode not present in any pure state, and the associated degeneracy may be lifted when higher-order corrections are accounted for. Since the equilibrium-HBH state is just a toy model with no microscopic underpinning, this issue seems unresolvable here. However, one may in any case expect the integral over $m$ to give prefactors in $e^S$, interpretable as logarithmic corrections to the entropy.

Let us now turn to a black hole in a cavity with a negative cosmological constant, in the case  $r_0\ll L$, where $L$ is the AdS radius.  As is clear from Eq.~\eqref{SHBHvsSBH}, any black hole of mass $M\leq r_0/3$ is entropically {\it favored} (see Fig.~\ref{fig:microcanonical}) to become a slightly cooler ball of NC gas that fills $r\leq 3M$, the region bounded by the original black hole's photon sphere.\footnote{Black holes with $M$ around and above $r_0/3$, including $M=r_0/3$, have positive specific heat, as normally defined. The fate of large AdS black holes in cavities will be addressed further in Paper III. }

Both here and in flat space, we see that a black hole of radius $\sim 2M$  can potentially evaporate into a cavity whose areal radius is $3M$. With only isotropic gas, such a process would be entropically disfavored; from \eqref{BH2isogas}, the cavity's radius would have to be parametrically larger.  If, in a realistic theory, black hole decay of this type were possible, it would motivate some reconsideration of the black hole information puzzle, since in this scenario  information stored in the black hole could not be carried off to great distances. We will return to this point  in Sec.~\ref{sec:meanings}.

All of this depends, of course, on the existence of the NC gas state and on its ability to persist around a black hole. Such issues cannot be addressed within this toy model, so we discuss them further in Sec.~\ref{sec:meanings}.

\subsection{Canonical Ensemble}\label{subsec:cavitycanon}

For completeness, we briefly discuss the canonical ensemble for an HBH in a cavity of radius $r_0$ held at fixed temperature. We begin with AdS space, where although the Hawking-Page transition remains unchanged by the existence of NC gas, subleading saddles are affected.

\begin{figure}
    \centering
    \includegraphics[width=0.75\linewidth]{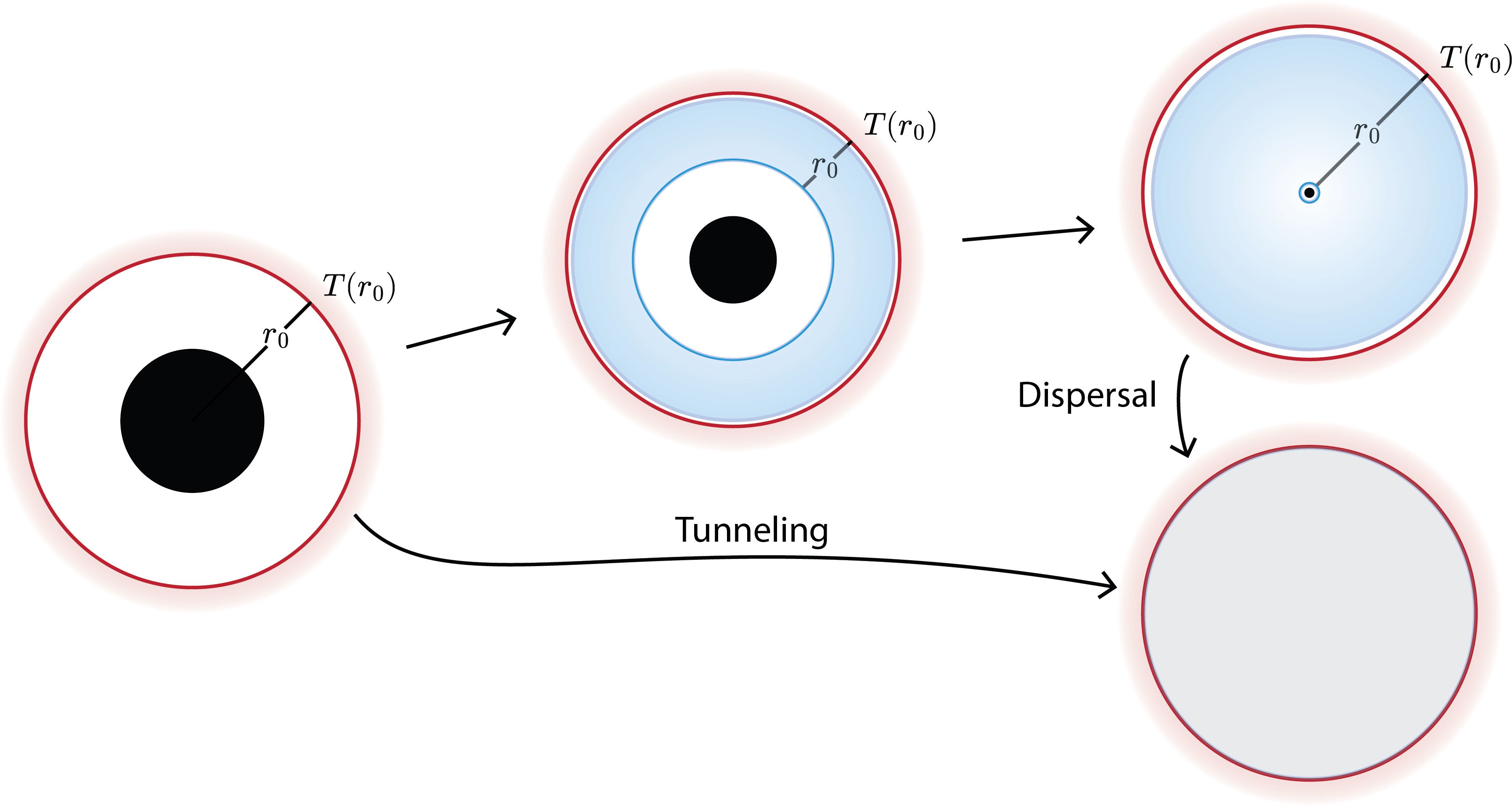}
    \caption{Evolution within the canonical ensemble for a (near-)spinodal black hole in a cavity with negative cosmological constant.  Rather than tunneling to the dominant state of diffuse ordinary gas, it may first become an HBH state, reaching a state of pure NC gas which then disperses to become the diffuse ordinary gas.}
    \label{fig:canonical}
\end{figure}

Black holes are topologically distinct from thermal AdS and are conventionally expected to tunnel to that state when it is dominant.  If the NC gas state exists, it may change this expectation. Small black holes and the spinodal black hole are subdominant to HBH states with the same temperature, including the NC gas state, to which they are continuously connected. This provides a smooth pathway  whereby small black holes, proceeding through intermediate HBH states as in Fig.~\ref{fig:canonical}, can form pure NC gas before dispersing to form the thermal AdS state; see also Paper III \cite{RiojasStrasslerAdS}.

 Large black holes near the spinodal temperature are subdominant to some HBH states and the NC gas state, which are in turn subdominant to thermal AdS. Each large black hole remains a local minimum of the free energy along an isotherm, even when globally subdominant. But instead of tunneling directly to thermal AdS -- a state with completely different topology -- a large black hole may tunnel through a small barrier to an HBH state with lower free energy.  From there it may smoothly evaporate to NC gas, and then disperse to form thermal AdS. 

Similar comments apply to black holes inside a cavity of radius $r_0$; see Fig.~\ref{fig:canonical}.  Details of the $r_0/L$ dependence are discussed in Paper III \cite{RiojasStrasslerAdS}, where York's flat-space analysis \cite{York:1986it} of a black hole in a cavity  is recovered in the limit $L\rightarrow \infty$.

As for the canonical ensemble in flat space, there the HBH and the pure NC gas have the same free energy as a black hole of mass $M$.  Since the free energy is degenerate, the transitions described above and shown in Fig.~\ref{fig:canonical} remain possible, but are no longer preferred by a free energy gradient. However, because the photon sphere is outside of the cavity on the large black hole branch, NC gas can form only on the small black hole branch (including the spinodal point).  Large black holes have lower free energy than all HBH states at that temperature, so those unstable to thermal flat space must tunnel to get there.

\section{Possible Connections to Black Hole Evolution}\label{sec:otherhints}

In this section we present two additional observations about the NC gas in equilibrium with a black hole.  First, we explore a formal property of the metric, which as $m\to 0$ can serve as a transition point between black hole metrics with horizons and thermal metrics with no horizon.  In the second, we follow Christensen and Fulling \cite{Christensen:1977jc} in exploring transverse pressure that is naturally created just below the photon sphere in the presence of Hawking radiation.

\subsection{A Geometric Transition at $m\to 0$}\label{subsec:conifold}

Imagine we begin with a black hole ($m=M$) and, keeping $M$ artificially fixed by sending in radiation to balance any Hawking radiation escaping to infinity, we allow or arrange for $m$ to decrease.  An ocean forms, and the black hole shrinks in size.  As $m\to 0$, the black hole disappears, leaving behind a spherical distribution of NC gas with zero radial pressure and a mild (and easily regulated) conical singularity at $r=0$ --- a traceless ocean with exactly the same entropy and 9/4 the surface area as the original black hole.

In the Gibbons-Hawking/York calculations of black hole thermodynamics, one compares the Euclidean metric of a black hole, whose radial sections are $S^1\times S^2$ with an $S^1$ shrinking to zero radius at the horizon $r=r_h$, to Euclidean flat space at finite temperature, whose radial sections are $S^1\times S^2$  with the $S^2$ shrinking to zero radius at $r=0$.  With this in mind, consider that (as noted in Paper I \cite{Riojas:2026ytt}) the Euclidean metric in the NC gas is a cone over $S^1\times S^2$, with metric 
\begin{equation}\label{metricoceanic}
ds^2=3dr^2 +r^2(\lambda_M^2 dt^2 + d\Omega^2)    
\end{equation}
The Euclidean metric of the HBH consists of three regions: a segment of the above cone in the region $3m<r<3M$ attached to inner and outer Schwarzschild metrics at smaller and larger $r$.

As $m$ decreases for fixed $M$, the cone extends toward smaller $r$. In the limit, the metric develops a conical singularity at $r=0$, in which both the $S^1$ and $S^2$ shrink simultaneously to zero size.  

From this singular metric one can then transition
to a new class of metrics,
in which the $S^1$ remains of finite size at the minimal value of $r$ while the $S^2$ shrinks to zero size there. (Recall spatially flat ${\mathbb  R}^3$ can be viewed as a trivial cone over $S^2$).  We can infer that 
the spherical traceless gas with a conical singularity
can potentially serve as a transition state between a black hole, where the $S^2$ remains of  finite radius at the horizon, and the thermal state of flat space where the $S^1$ remains of finite size.   This is reminiscent of the conifold transition (in which an $S^2$ and $S^3$ play similar roles); though a formal, kinematic property of metrics, this transition has been shown in some cases to play a role in actual dynamics.

The same logic applies in any spacetime dimension $d>4$.  The Euclidean metric is a cone over $S^1\times S^{d-2}$, which serves as a transition metric between metrics with a black hole horizon and thermal metrics without a horizon. 

The above observations may be applied to a black hole surrounded by any self-similar  extended photon sphere ($\delta=2)$, even if misaligned ($\nu\neq 1/3$).  The fluid region then has a metric  of the form and topology of \eqref{metricoceanic} but with $g_{rr}=1-2\nu$.   However, misaligned extended photon spheres have non-zero radial pressure and Ricci scalar, and thus require walls or transition regions at their edges.  The importance of this complication deserves further study; for some discussion see \cite{Riojas:2026god}.

\subsection{Christensen-Fulling Perspective on Ocean Formation}\label{subsec:CF}

Our HBH calculations would be that much more interesting if an \hillingas\ could  form naturally in the course of black hole evaporation.  Though evidence in favor of this possibility is very thin, there is an intriguing suggestion to be found in old work of Christensen and Fulling (CF) \cite{Christensen:1977jc}. 

CF considered quantum fields in the background of a Schwarzschild solution, and showed that conservation of their stress tensor ${\cal T}_{\mu\nu}$, along with the trace anomaly caused by the curved background, leads to a relation between the outward flux of energy, a radial integral of the trace anomaly integrated across space, and a radial integral of the transverse pressure ({\it i.e.} the angular components of ${\cal T}_{\mu\nu}$.)  From this relation, they showed that a field contributing to Hawking radiation will develop, as a result of the trace anomaly, either positive transverse pressure for $r$ below the photon sphere or negative transverse pressure for $r$ above the photon sphere. 

Although they could not distinguish between these two possibilities, CF gave an argument in favor of the former, speculating  that positive transverse pressure below the photon sphere is associated with the instability of photon sphere geodesics and wave modes. Recalling that Hawking radiation to $r=\infty$ is the tail of Rindler radiation near the black hole, which includes modes of all angular momentum $\ell$, they suggested that modes with $\ell>0$ and moderate energy 
might become trapped just below the photon sphere, with some remaining there for an exceptionally long time.   From this point of view, the presence of these modes, an inevitable consequence of the mechanism that generates Hawking radiation, might be responsible for the transverse pressure. 
This result hints that quantum effects might cause the spontaneous generation of an ocean, beginning at the black hole's photon sphere and growing downward as the black hole evaporates.
 While we are not able to check this idea directly,  we are able to strengthen the CF analysis.  In Appendix~\ref{app:CF2}, where we review CF's argument and repeat it for an HBH. we show that transverse pressure is always generated {\it below the ocean's lower surface} at $r=3m$, not at its upper surface at $r=3M$.  In CF, distinguishing between the regions above and below the photon sphere was not possible, because $m$ and $M$ are equal for a black hole and are thus conflated. But with a photon ocean, the $m$- and $M$-dependence can be parametrically separated, most unambiguously when $m\ll M$. 

 In short, the combination of Hawking radiation and the trace anomaly for a typical quantum field will cause the field to generate positive transverse pressure $P_\perp \sim \hbar / r^4\sim \hbar/m^4$ in the region just below the HBH ocean.  This is consistent with CF's suggestion concerning the cause of the transverse pressure.
 Whether this transverse pressure can build up over time to $P_\perp\sim 1/Gr^2$, enough [see Eq.~\eqref{gassmassfunction}] to cause back-reaction on the metric and allow the ocean of NC gas to deepen over time, is a question that requires more sophisticated methods.

\section{Discussion}\label{sec:meanings}

In Paper I \cite{Riojas:2026ytt}, we explored classical properties of the HBH metrics, solutions to the Einstein equations coupled to an ocean of NC gas encoded in a stress-energy tensor. In this paper we have pushed these solutions to the edge of physical plausibility, by demanding that the black hole and the surrounding ocean be in thermodynamic equilibrium and computing the consequences.  We have taken the view that this makes the HBH  a toy model for a black hole in thermal quasi-equilibrium with surrounding radiation.  As we do with all toy models, we must ask if such a model captures features of a realistic system.

Of course the toy model itself cannot answer the question.  Until we have a concrete realization of an analogous system in a full quantum theory, or a proof that such realizations are impossible, we can only speculate.  Our goal in this final section is to explore the issues, noting strong reasons for pessimism that are counter-balanced by the uniqueness of the model and its striking features.  Even if one takes a maximally pessimistic view, we would argue that the model raises enough questions that it merits further exploration.

\subsection{Uniqueness of the Model}\label{subsec:HBHIsUnique}

At the heart of the equilibrium-HBH model is the assumed existence of an ocean of cold, equilibrium NC gas. This state 
has many special properties.  Among self-similar spherically symmetric solutions with a constant ratio of pressures to density, it has pride of place, appearing at the intersection of four interesting curves as shown in Fig.~\ref{fig:nudelta}. 
{Located on the curve of traceless fluids, the NC gas can be made from excitations of massless fields, such as those dominantly emitted in Hawking radiation; with $P_r=0$, it requires no walls to contain it or edge regions to spoil its self-similarity; with $\delta=2$, it forms an extended photon sphere and thus has $S\sim M^2$, making it conceivable that it might carry the same entropy as a black hole; 
and with $\nu=\frac13$, its extended photon sphere is aligned, so that it can be sourced by massless quanta orbiting a central black hole.  Only for the NC gas are these four possibilities  simultaneously true.}

The NC gas solutions remain similarly special for any dimension above four.  Moreover, among the self-similar solutions  with $\widehat m\propto r^{d-3}$, the NC gas is the only one that also exists in AdS and dS spaces (see Paper I \cite{Riojas:2026ytt}, Sec.~3.).

The NC gas metric, an extended photon sphere, is interesting geometrically.  It  everywhere satisfies the condition $rf'/2f=1$ that characterizes a photon sphere, and as seen in Sec.~\ref{subsec:conifold}, its Euclidean version is a section of a cone over $S^1\times S^2$.  Because this extended photon sphere is aligned, its upper and lower edges join to Schwarzschild metrics without need for any thin (distributional) layer of stress-energy at the junctions, or for any transition regions.

The equilibrium-HBH solution demands the black hole of mass $m$ and the NC gas have the same temperature.   It is non-trivial that this is automatic for the NC gas; in flat space, Eq.~\ref{THLargeR}, combined with the Tolman redshift of temperatures, assures this is so. In Appendix \ref{app:coldestNC} we show that the self-similar $d=4$ NC gas solution is colder than any other combination of smooth and singular arrangements of NC gas.  In \cite{Riojas:2026god} this is non-trivially generalized to all possible mechanically-stable Einstein clusters, a point deeply related to the black hole's heat capacity [see Eq.~\eqref{specificheat}] and the fact that it changes sign at the photon sphere.  Thus all other Einstein clusters that can be mechanically stable are hotter than a black hole, and cannot be in equilibrium. 

Furthermore, the NC gas is metastable in the following sense shown in \cite{Riojas:2026god}. Thin shells of material placed around a black hole will generally be either mechanically unstable or thermodynamically stable, unless they have vanishing mass, are traceless, and sit at the photon sphere \cite{Brady:1991np}. The NC gas can be understood as a series of infinitesimally light traceless shells with finite total mass. This is because self-gravitating gases at zero radial pressure form shells satisfying the IJCs \cite{Riojas:2026god}. Therefore the NC gas uniquely satisfies (marginal) mechanical and thermodynamic stability while surrounding a Schwarzschild black hole. 

It is that much more remarkable, then, that the resulting HBH in asymptotically flat space is not only an optical mimic (see Paper I \cite{Riojas:2026ytt}) of a black hole of mass $M$ but also, under the assumption of equilibrium, a thermodynamic mimic.\footnote{Note this is true of overall thermodynamic quantities $(T,S,F)$, but presumably does not extend to gray-body factors and other subleading effects.} Mimicry follows directly from properties of the NC gas: its bulk  action vanishes because the gas is traceless, the matter action vanishes because the radial pressure is zero, and there are no wall contributions, leaving the Euclidean action calculation the same as for a black hole.  Said another way, the NC gas satisfies $\rho = 2P_\perp = sT$, $P_r=0$, and direct calculation shows that a mimicking temperature leads to mimicking entropy.

That said, it is shown in \cite{Riojas:2026god} that conditions underlying thermodynamic mimicry strongly resemble those allowing for extended photon spheres. When appropriate massless walls are placed at the edges of a misaligned extended photon sphere, thermodynamic mimicry occurs, with  $S = 4 \pi M^2$ and $T = (8 \pi M)^{-1}$.  This observation applies to  any member of the $\delta=2$ family of self-similar solutions in Fig.~\ref{fig:nudelta}, including ``frozen stars" \cite{Brustein:2018web,Brustein:2021lnr,Brustein:2023hic} and ``stiffest stars" \cite{Banks:2002fj} as well as the HBH. 
However, massless walls violate the dominant energy condition. For the extended photon sphere of the HBH, which is aligned and requires no walls, there is no risk of violating energy conditions and the thermodynamic mimicry is more straightforward.

Considering the mimicry, the shell-by-shell metastability, the relation to Hawking radiation, and the uniqueness, one has to wonder why the equilibrium-HBH model exists and what its role in physics might be.
It may be that the answer is formal rather than physical, and that the model is a clue to some mathematical or conceptual property of semi-classical or quantum gravity. Perhaps we may interpret the results of \cite{Riojas:2026god} as reflecting certain mathematical transformations in semiclassical gravity and its thermodynamics, akin to the one discussed in Sec.~\ref{subsec:YorkInterpret}, in which a black hole can be  transformed into another system of bulk material with the same thermodynamic behavior.  But in any case, it seems highly improbable to us that the mathematics of general relativity would offer us this model, with all of its remarkable properties, if it had no theoretical significance of any kind. 

\subsection{Challenges and Possible Opportunities}\label{subsec:maxpess}

As promised at the start of this paper, we now consider various arguments that suggest one should regard the HBH system, when placed in thermal equilibrium with a black hole, as a toy model.   At a bare minimum, one must be wary in interpreting its behavior.   Nevertheless, history has examples of toy models that captured an essential aspect of a realistic physical system for the wrong reasons.  We therefore think it too early to discard the model, and in this spirit, we now cautiously explore its limitations and possibilities.

An immediate question is whether the state of cold NC gas, required by the model, can exist in any complete theory.  At such a low temperature $T(r)\sim \hbar/r$, the wavelengths of the gas quanta are comparable to the ocean radius; treating it as a region of non-zero $T_{\mu\nu}$ with sharp edges at $r=3m$ and $r=3M$ is unjustifiable.  Moreover, at these temperatures, the NC gas has enormous entropy density $s\sim 1/\hbar G r$, and it is difficult to see how to concretely build such a state out of microscopic, quantum field theoretic ingredients.\footnote{We have treated the NC gas as a perfect anisotropic fluid, which suggests a simple picture of massless particles executing circular orbits, with a traceless stress-energy tensor. While this holds in the geometric optics limit, exact tracelessness in field theory demands Weyl invariance of the matter action. This feature is satisfied by massless Dirac fermions, and by Maxwell fields in $d=4$, but massless scalars are Weyl invariant only when conformally coupled to gravity.}

In the language of 't Hooft $N_c$-counting,  this large bulk entropy density would imply a field theory entropy density of order $N_c^2$.  Such an entropy density is characteristic of the deconfined CFT phase dual to a black hole horizon in AdS/CFT, not of the confined CFT phase that is dual to a gas of massless bulk quanta.  We might therefore expect a cold NC gas to  be dual to an unfamiliar, high-entropy confined state, and a large HBH to be dual to a two-phase system.  In AdS/CFT, this high-entropy state might only exist for a CFT (or other QFT) at strong coupling on a sphere, a context where confinement is kinematic \cite{Aharony:2003sx,Aharony:2005bq} and is still poorly understood.  We will explore this further in Paper III \cite{RiojasStrasslerAdS}.

It is also possible that the required state is not, in fact, a gas of massless excitations orbiting the photon sphere of the black hole, and is instead an exotic fluid that happens to share the same equation of state.  This might make the model more consistent but also might reduce its appeal, as it could disconnect it from the massless excitations that naturally appear in Hawking radiation.

Note also that although the gas has a species problem, it is exactly the same one found in the context of black holes and Hawking radiation.
(For simplicity of discussion, let's assume here  that all species are either massless or at the Planck scale.) Specifically, one would normally expect gas thermodynamics to depend on the cocktail that makes up the gas: what fraction is gravitons, what fraction is photons of one or another species, etc.  The toy model subsumes the makeup of the NC gas into $T_{\mu}^{\ \nu}$, hiding any dependence on the specific cocktail.  This might seem inconsistent, but the thermodynamics of the gas, like that of the black hole, is calculated in classical gravity, and it is well-known  that species-dependence in black hole calculations enters only at loop-level \cite{Susskind:1994sm}. Loops of massless particles, all of which appear equally due to the universality of gravity, both renormalize $G$ and ensure that all massless species in the theory are found in the Rindler atmosphere and in the Hawking radiation. 
Here we might expect the situation to be the same: the gas entropy density $s\sim 1/G$ should be calculated with the renormalized $G$, and the ocean should contain the same cocktail of massless species found in the Rindler atmosphere and Hawking radiation.  Indeed, if the latter were false, then one might wonder how the black hole and ocean could be in equilibrium: exchange of modes between the ocean and the Rindler atmosphere and/or Hawking radiation would alter the ocean's cocktail.\footnote{Note this species problem differs somewhat from the one studied in, for example \cite{Dvali:2007hz,Dvali:2007wp}, where one explores the effects of taking $N_s$ large on a gas where $\rho,s$ depend explicitly on $N_s$.  In our system $N_s$ may be small, and also $N_s$ does not appear explicitly in $\rho, s, T$, which are all fixed by gravity and thermodynamics. Instead $N_s$ can enter only through the renormalization of $G$, which appears in $\rho\sim 1/Gr^2$, $s\sim 1/\hbar Gr$ but not in the local temperature of the NC gas, $T(r)\sim \hbar/r$.}

Even if the ocean of cold NC gas can exist, establishing and maintaining thermal equilibrium with a black hole is questionable.  Although the NC gas around a black hole is more stable thermodynamically and mechanically than other similar systems \cite{Riojas:2026god}, important issues remain. First, the ocean has unstable geodesics just outside its upper and lower surfaces, the same as those at any photon sphere. It is thus potentially subject to rapid leakage, which could drain the ocean before it can reach equilibrium.  Second, there is a potential inconsistency:  thermal equilibrium normally requires interactions, but a $P_r=0$ fluid is destabilized by interactions. And third, the negative specific heat of the black hole can potentially pull the ocean and black hole out of equilibrium.

Regarding these issues, it is useful first to ask how a black hole and its ocean, and a possible heat bath, could ever be in equilibrium if interactions in the NC gas are rare.  Consider a gas of photons inside a gigantic hollow blackbody; the photons will have a thermal spectrum, despite the rarity of interactions, due to the sources and sinks of those photons being at a definite temperature.  We might therefore attempt to interpret the temperature of the ocean in a similar way. In the canonical ensemble the dominant source/sink would be the heat bath, though it is not obvious how a bath could populate a gas in a $P_r=0$ state.  In the microcanonical ensemble the source and sink would have to be the black hole and its radiation.  In reviewing and extending the results of Christensen and Fulling \cite{Christensen:1977jc} in Sec.~\ref{subsec:CF}, we saw that transverse pressure is found below the lower edge of the ocean.  This may be due (as suggested in \cite{Christensen:1977jc}) to $\ell>0$ modes of the Rindler atmosphere that almost escape the black hole and spend an extended period orbiting near the photon sphere.  If leakage in and out of the ocean's lower surface could allow the black hole and ocean to exchange modes, it would not be surprising that the temperatures match.  

However, the last remark begs the question: are the time scales consistent?  A black hole of mass $M$ evaporates slowly: it emits one quantum to infinity, with energy $T\sim 1/{GM}$, in a time $t_0\sim GM$, the black hole crossing time.  It emits $m_{pl}$ of energy to infinity over a time $\sim M^2$, and evaporates in a time $\sim M^3$.  Meanwhile the ocean's leakage time scale is naturally the inverse Lyapunov exponent \eqref{Lyapunovs}, $\lambda\sim GM$, times a logarithm.  In any imaginable cold NC gas, whose quanta have wavelength $\sim1/T\sim GM$, it is not easy to see how leakage out of the ocean would be slow enough to be balanced by emission from the black hole.

Finally, the equilibrium-HBH model implies that, in the microcanonical ensemble, all black holes in AdS spaces, including large ones, will decay to cold NC gas. This is certainly not expected, though there are subtleties and loopholes associated to this statement --- see Paper III \cite{RiojasStrasslerAdS} --- so we remain agnostic about its meaning.  In flat space, black hole decay to NC gas is entropically allowed but not required, since (at the leading order) the model predicts degeneracy between HBH states of the same $M$.

With this in mind, let us return to what the model might suggest about black hole evolution. We considered black holes in small cavities, with radius $r_0\gtrsim 3M$, in Sec.~\ref{sec:cavity}. In the canonical ensemble, only subleading saddle points in AdS are significantly affected. But in the microcanonical ensemble something more interesting happens: black holes which would be expected to evaporate only partially can now evaporate completely if NC gas is available.  Specifically, if the cavity has no cosmological constant, the spinodal black hole with $M=r_0/3$ {\it may} evaporate, as its entropy is degenerate (within the toy model) with that of a black hole of the same mass. If the cavity has a negative cosmological constant, the black hole with $M=r_0/3$ {\it will} evaporate, as it has less entropy than the gas state.  The metric transition of Sec.~\ref{subsec:conifold} would then be realized, at least approximately, within the cavity.

This possibility would add elegance to the story of flat-space black holes.  The power of 5/3 in \eqref{BH2isogas} seems rather ungainly, and makes the canonical and microcanonical ensembles very different.  But with the NC gas state available, the spinodal black hole in a cavity of radius $r_0$, with zero inverse-heat-capacity and with its photon sphere at the cavity's edge, becomes an important feature of both ensembles. This would be true  in all dimensions $d\geq 4$. 

If a black hole could actually evolve into gas within a small cavity, as the toy model suggests, this would be conceptually important, and potentially in tension with usual expectations in certain settings. For example, in the pedagogical lectures of Mathur  \cite{Mathur:2009hf} on black holes and the reasons to consider fuzzballs \cite{Mathur:2002ie,Mathur:2008nj}, a model of bits is used to argue that the fine-grained entropy must increase with each step in Hawking radiation as long as certain conditions are satisfied. One such condition, used repeatedly in the argument, is that of locality; because Hawking bits emitted long ago are far away, locality prevents them from interacting with those forming at the present. From this it is argued that the entanglement entropy of the set of previous bits does not change when a new bit forms, a fact essential to the proof, which involves computing entropies of unions of subsets. Mathur's bit model is also used in the discussion by Polchinski \cite{Polchinski:2016hrw} of complementarity and firewalls \cite{Almheiri:2012rt,Almheiri:2013hfa}, notably in the context of potential  violations of locality. 

But in a setting where a black hole in a cavity evolves into an HBH, its earlier Hawking modes would be stored nearby, so that the locality argument would not hold. 
Indeed, given the low temperature and long wavelength of the Hawking quanta, the black hole and the ocean would presumably act as an interacting system that cannot be separated into two independent parts. It might not be possible to view any modes that they exchange as on-shell, undermining the foundation of Mathur's bit model. 

Thus,  if the equilibrium-HBH model captures something true about evolution in a cavity, certain arguments that play a role in our current understanding of black holes would need extension or revision.  Whether it would affect our fundamental picture of evaporation and information\footnote{Major progress in quantum gravity has substantially addressed the information problem \cite{Penington:2019npb,Almheiri:2019psf,Almheiri:2019yqk,Penington:2019kki,Almheiri:2019hni,Almheiri:2019qdq}. There remain some open questions, especially in higher dimensions, see \cite{Raju:2020smc,Giddings:2020yes,Geng:2020qvw,Geng:2020fxl,Geng:2021hlu,Antonini:2025sur,Geng:2026asi}. If islands form outside the stretched horizon \cite{Matsuo:2020ypv,Bousso:2023kdj}, it would be interesting to consider the HBH, and some formalism in \cite{Riojas:2026god}, in that context. In doubly-holographic models, islands on the brane are sensitive to photon spheres, see \cite{Karch:2023ekf}.} is far from clear. 

We conclude our speculations with one that strains credulity to the breaking point: if a black hole in a cavity could evaporate with the ocean of NC gas playing a central role, is there any chance that an ocean of NC gas could be important even for black holes evaporating {\it into infinite space?} If an ocean forms naturally during the evaporation process, the local temperature at the ocean's surface (in asymptotically flat space) matches that of Hawking modes, so an observer collecting excitations at infinity cannot easily determine their origin from temperature alone.  As shown in Sec.~\ref{sec:AdS}, the model  predicts that  a small AdS black hole is entropically favored to become an HBH, and eventually a ball of NC gas.  Also, in Sec.~\ref{subsec:CF} following \cite{Christensen:1977jc}, we saw that for a black hole emitting Hawking radiation to infinity, transverse pressure arises below the ocean's lower surface, a prerequisite for the ocean to deepen.    Are these hints meaningful or misleading?

It must be stressed that ocean formation is significantly less plausible in infinite space than in a small cavity. With a cavity wall at the upper surface of the ocean, leakage from the ocean's upper surface is inhibited.  Furthermore, the ADM mass $M$ of the black hole and ocean are constant in the cavity, even if the black hole mass $m(t)$ is decreasing over time. In infinite space, the ADM mass of the black hole and ocean\footnote{This should be measured at a distance $R$ with $M\ll R\ll M^2$: far enough to allow an ADM-like measurement, but close enough that most Hawking radiation is not included in that measurement.} will no longer be constant.\footnote{This notion suggests amusing scenarios. For instance, suppose a black hole could populate a surrounding  ocean at a rate comparable to the (local) rate at which it emits Hawking radiation.  This would cause the black hole mass $m(t)$ to decrease faster than the HBHs ADM mass $M(t)$.  An observer at infinity, wanting to test the firewall argument \cite{} but failing to notice the subtle difference between a black hole of mass $M(t)$ and an mimicking HBH of the same mass, would overestimate the Page time $t_{Page}$. Descending after $t_{Page}$ to small $r$, in order to explore how late-time modes are entangled with those inside the black hole, the observer would be shocked to discover that the black hole is already gone.}  With both $M(t)$ and $m(t)$ decreasing, there is a risk that the ocean may find itself outside the system's photon sphere and may consequently disperse.  Dynamics of a rather magical sort might be needed to prevent this from being a generic outcome.

Nevertheless, it is important to rule out this possibility  definitively.  Just as in the cavity, an ocean storing some of the early-time radiation and information in the vicinity of the black hole would allow late-time and  early-time radiation to interact, and would make it difficult to identify late-time Hawking modes until they had already passed into or through the ocean.  These would challenge the common assumptions that the entanglement of early-time radiation cannot change due to later radiation, and that the entanglement of individual Hawking modes can be straightforwardly determined.

We should emphasize that we have found no evidence, other than the hints already mentioned,  that an ocean might form naturally as an isolated black hole in flat space evaporates. But given the potential consequences, we view it as important to exclude this scenario. This will require exploring complete quantum theories with the equilibrium-HBH model and its cold NC gas in mind.

\subsection{Final Remarks}

Looking ahead, there are various directions to pursue.  Most important would be the identification of the state in a quantum theory corresponding to the NC gas.  But simpler opportunities are also in view.

It would be interesting to find a lower-dimensional generalization of an HBH.
The solutions found above require $d\geq 4$ and become nonsensical in $d=3$.  Nevertheless, NC gas itself is not excluded in $d=3$, so perhaps a generalization of the HBH exists.  Since such solutions would be dual to $d=2$ field theories, they might offer a concrete setting where the quantum aspects of HBHs could be further studied.

Additional insights might be gleaned by generalizing the HBH to charged, rotating and perhaps supersymmetric black holes. In rotating black holes, there is a photon zone rather than a single photon sphere, and an ansatz for what the rotating HBH should be is not obvious.  Furthermore, it is not clear whether one can or should treat the black hole's angular momentum and that of the surrounding gas as independent.  Complications arise also for  Reissner-Nordström black holes: if the gas carries charge, then its behavior depends on the charge-to-energy ratio of its constituents, since the locations of stable orbits depend on that ratio.  While we see indications that such solutions should exist, the corresponding TOV equations are more complex, and  the best ansatz is not as obvious as in the Schwarzschild case.  It may nevertheless be straightforward.

We also emphasize that if HBHs can be shown not to exist --- if NC gas can be excluded or shown to be highly unstable --- then an important loophole would be eliminated. In \cite{Brady:1991np}, obstructions were posed to black holes reaching equilibrium with surrounding fluid shells, due to mechanical and thermodynamic instabilities that could not be evaded, with one apparently trivial exception.  The equilibrium HBH would generalize that exception \cite{Riojas:2026god}. Therefore, if HBHs cannot in fact reach equilibrium, this would close the loophole in \cite{Brady:1991np}, and a rigorous ``no go'' theorem might be possible.

To conclude, let us summarize our perspective based on this paper, on Paper I \cite{Riojas:2026ytt}, and on \cite{Riojas:2026god}. Additional results in the AdS context will follow in Paper III.  

The equilibrium-HBH model of a black hole in thermal equilibrium with an ocean of NC gas appears to be rare and perhaps unique, both in the properties of the gas itself (Fig.~\ref{fig:nudelta}) and in its mechanical and thermodynamic stability around a black hole \cite{Riojas:2026god}. It provides us with a system, in all dimensions $d\geq 4$, that is both an optical and a thermodynamic mimic of a black hole in flat space, and suggests the possibility that black holes could evaporate in small volumes.  Its simple and elegant results seem worthy of some attention.   That said, there are numerous reasons to think that in a realistic theory, an equilibrium HBH state with its enormous entropy density and potentially leaky ocean cannot occur.  Perhaps, instead of the model capturing some essential physics of real black holes, it points us toward unknown general relations between black holes and bulk materials built from matter shells, such as are noted in Sec.~\ref{subsec:YorkInterpret} and in \cite{Riojas:2026god}.  One might wonder if there is something holographic in this relation, though we see no sign of this in the AdS context, where it would be most naturally expected.  

Yet certain special properties of the NC gas --- its conical metric over $S^1\times S^2$, its connection with the photon sphere and the trace anomaly, and its potentially significant impact on black holes in cavities --- suggest that it is too early to abandon the idea that HBH states might have a role to play in black hole evolution.  Though the possibility is at best remote, its implications would be substantial. We therefore view it as worth exploring further, with an appropriate level of skepticism.

\begin{acknowledgments}
     We thank Håkan Andréasson, Andreas Karch, Akshay Ghalsasi, Daniel Jafferis, Alex Lupsasca, Rashmish Mishra, Sonia Paban,  Suvrat Raju, Lisa Randall, Mukund Rangamani, Matt Reece, Andrew Strominger, Hongji Wei and Lawrence Yaffe for helpful comments and conversations. Calculations used the Mathematica packages diffgeo.m (Headrick) \cite{Headrick:diffgeo} and OGRe \cite{Shoshany:2021iuc} (Shoshany). MR is supported by the Gravity, Spacetime, and Particle Physics (GRASP) Initiative at Harvard University. MJS thanks Harvard University's Department of Physics for its long-term hospitality.
\end{acknowledgments}

\bibliographystyle{JHEP}
\bibliography{biblio}

\appendix

\section{The HBH for $\Lambda \ne 0$ and Higher Dimensional Spaces}\label{app:otherHBH}

As shown below, HBH solutions exist in all $d\geq 4$ with any cosmological constant, and have similar unique properties in each dimension.   For other choices of $P_i/\rho$, linear self-similar solutions exist only if $\Lambda=0$; some of these self-similar solutions are given in Paper I.

Define 
\begin{equation}\label{eq:Acd}
  \Omega_{d-2}=     \frac{2 \pi ^{\frac{d-1}{2}}}{\Gamma \left(\frac{d-1}{2}\right)}  \ \ \ \ ; \ \ \ \
\gamma_d=\frac{16 \pi G }{{(d-2) \Omega_{d-2}}{}} \ .
\end{equation}
For a black hole of mass $M$, its horizon radius $r_{h,M}$ and the radius of its photon sphere $r_{\mathrm{ps},M}$ are 
\begin{equation}
r_{h,M} = \left(\gamma_d M\right)^{1/(d-3)} \ \ , \ \ 
r_{\mathrm{ps},M}= \left(\frac{d-1}{2}\right)^{{1}/{(d-3)}} r_{h,M}
\ .
\end{equation}
Define also
\begin{eqnarray}
\tilde\lambda_d&=&
\sqrt{\frac{  d-3}{d-1}} \left(\frac{2}{(d-1) }\frac{1}{\gamma_d M  }\right)^{\frac{1}{d-3}}
=\sqrt{\frac{(d - 3)}{(d - 1)}}\frac{1}{r_{\mathrm{ps},M}} \ ,
\end{eqnarray}
which differs from the usual Lyapunov exponents by a factor of $\sqrt{d-3}$.
Then 
the HBH metric takes the form
\begin{equation}
    ds^2=-f(r) dt^2 + j(r)^{-1} dr^2 + r^2 d\Omega_{d-2}^2
\end{equation}
with
\begin{equation}
    j(r) = 1- \frac{\gamma_d \hat m(r)}{r^{d-3}} \ \ ; \ \ f(r)=\left[\frac{\tilde\lambda_{d}}{\hat\lambda}\right]^{2} j(r)
\end{equation}
and
\begin{eqnarray}
\hat m(r) =
\begin{cases}
m \ ,&   r\in(r_{h,m},\,r_{\mathrm{ps},m}),\\[2pt]
\frac{2}{d-1}\frac{r^{d-3}}{\gamma_d}  \ ,&   r\in(r_{\mathrm{ps},m},\,r_{\mathrm{ps},M}),\\[2pt]
M\ ,&  r \in(r_{\mathrm{ps},M},\,\infty).\\[8pt]
\end{cases}
\end{eqnarray}
Here  $\hat\lambda$ takes the same form as $\tilde\lambda_d$ with $M$ replaced with $\hat m(r)$. Note that $f(r)=\tilde\lambda_d^2 r^2$  and $j(r) = \hat\lambda^2 r^2$ in the ocean.  The ocean metric as $m\to 0$ is a cone over $S^1\times S^{d-2}$. 

The generalization to AdS spaces is straightforward. In AdS$_4$, with $\Lambda=-3/L^2$,  the  HBH metric takes the form
\begin{align}\label{HBHAdSmetric0}
    j(r)\equiv a^{-1} = 1+\frac{r^2}{L^2}-\frac{2\widehat{m}(r)}{r}  \quad , \quad \quad 
    f(r) 
    = \left[\frac{\lambda_M}{\hat \lambda(r)}\right]^2  j(r) .
\end{align}
with the fixed and variable Lyapunov exponents  
\begin{equation}
    \lambda_M^2 = \frac{1}{27M^2} + \frac{1}{L^2}  \ ;
      \quad \quad 
    \hat\lambda_{ }^2 = \frac{1}{27[\widehat{m}(r)]^2} + \frac{1}{L^2}
    \ ,
\end{equation}
and the familiar mass function
\begin{equation}\label{massfunctionAdS}
\widehat{m}(r)=
\begin{cases}
m, & r\in(r_+,\,3m),\\[2pt]
\dfrac{r}{3}, & r\in(3m,\,3M),\\[2pt]
M, & r\in(3M,\,\infty).
\end{cases}
\end{equation}
Here the central black hole's horizon  $r=r_+$ satisfies
\begin{equation}\label{rplus}
    1-\frac{2m}{r_+}+\frac{r_+^2}{L^2}= 0
\end{equation}

This solution is easily generalized to AdS$_d$ space for $d>4$.  Equations \eqref{HBHAdSmetric0}--\eqref{massfunctionAdS} apply, but with $2\widehat{m}$ replaced with $\gamma_d\widehat{m}$ [where $\gamma_d$ is defined in Eq.~\eqref{eq:Acd}]. The mass function $\widehat{m}$ in the ocean is
\begin{equation}
    \widehat{m}(r)= \frac{2 }{(d-1)\gamma_d}r^{d-3}
\end{equation}
and  the quantities 
\begin{equation}\label{Lyapunovd}
    \tilde\lambda_{d}=\sqrt{
{\frac{  d-3}{d-1}} \left(\frac{2}{(d-1) }\frac{1}{\gamma_d M  }\right)^{\frac{2}{d-3}}+\frac{1}{L^2}} \ \ \ ; \ \ \
    \hat \lambda=\sqrt{
{\frac{  d-3}{d-1}} \frac{1}{r^2}+\frac{1}{L^2}} \ \ . \ \ 
\end{equation}
which differ from the usual Lyapunov exponents by a factor of $\sqrt{d-3}$.
The horizon $r_+$ now satisfies $1+r_+^2/L^2-\gamma_dm/r_+^{d-3}=0$.

All the above metrics involve solutions to the $d\geq4$ TOV equation, which we now provide. For a spherically symmetric metric analogous to \eqref{metricform}, we have (including a cosmological constant $\Lambda \equiv -\frac12(d-1)(d-2)L^{-2}$)
\begin{equation}
    j(r) = 1- \frac{\gamma_d\widehat{m}(r) }{r^{d-3}}-\frac{2 \Lambda r^2}{(d-1)(d-2)}
 \end{equation}
The TOV equation then reads 
\begin{equation}\label{TOVd}
    P_r^{\prime}(r)=-\frac{\rho(r)+P_r(r) }{r j(r)}
    \left(\frac{\gamma_d}{2}
    \Omega_{d-2} r^2 P_r(r)+\dfrac{r}{2}\dfrac{\partial j(r)}{\partial r}
    \Bigg|_{\widehat{m}}
   \right)
    -\frac{(d-2)}{r}\left(P_r(r)-P_{\perp}(r)\right) .
\end{equation}

Einstein clusters and self-similar solutions are present for any $d>4$; see Paper I for details.  Here we note self-similar solutions with no central black hole take the form
\begin{equation}\label{massfunctiondx}
\widehat{m}(r)= \frac{2 }{(d-3)\gamma_d}\nu r^{d-3} \ \ \ \Rightarrow \ \ \ \    \rho = \frac{\widehat{m}'(r)}{\Omega_{d-2}r^{d-2}}=\frac{(d-2)\nu}{8\pi r^2} \ .
\end{equation}
Define
\begin{eqnarray}
    w_x=2w_r +(d-2)(w_\perp-w_r) &=&(d-2)w_\perp-(d-4)w_r
\end{eqnarray} 
which equals 1 for the NC gas.
Then inside the gas
\begin{eqnarray}\label{eq:fjhigherd}
    \nu &=&\frac{w_x}{\frac{2 w_x}{d-3}+(1+w_r)^2} \ \ ; \ \ 
    \delta =\frac{2\nu(1+w_r)}{j}=\frac{2 w_x}{1+w_r}
\nonumber \\
    j_0\equiv j(r)&=&1-\frac{2\nu}{d-3}=1-\frac{\frac{2 w_x}{(d-3)}}{ \left(\frac{2 w_x}{d-3}+(1+w_r)^2\right)} \ \ ; \ \ f(r)\propto r^{\delta}
\end{eqnarray}
The $\delta$ dependence of $f(r)$ is of particular importance.

In Paper I we also considered self-similar  $d=4$ Einstein clusters around a black hole, with $w_r=0$, $\nu=2w_\perp /(1+4w_\perp)\leq \frac13$, and $\delta=4w_\perp= 2\nu/(1-2\nu)$.  The metric generalizes the HBH solution for $\nu=\frac13$:
\begin{align}\label{jhnulllinearmassBH}
    j(r)= 1
    -\frac{2\widehat{m}(r)}{r}  \quad , \quad \quad 
    f(r)= \left[\frac{\widehat{m}(r)}{M}\right]^\delta  j(r) 
\end{align}
and 
\begin{equation}\label{massfunctionlinearBH}
\widehat{m}(r)=
\begin{cases}
m, &   r\in(2m,m/\nu)\\[8pt]
\nu r,  &   r\in(m/\nu,M/\nu)\\[8pt]
M , &   r\in(M/\nu,\infty)
\end{cases} \ \ \\ 
\end{equation}
Solutions of this type have appeared as a special case of Model 1 of \cite{Maeda:2024tsg}. 

\section{Scaling Laws for Fluids}\label{app:scaling}

Scaling laws for fluids in $d\geq 4$, extending those of Sec.~\ref{subsec:detailedpowerlaws}, are easily found.
In general $d$, we have $r_0^{d-3}\sim M$ for a relativistic gravitating system, while its surface area scales as (but is somewhat larger than)  the horizon area of a black hole in $d$ dimensions:
\begin{equation}
    A_{d-2}[M]\sim r_0^{d-2}\sim M^{(d-2)/(d-3)} \ .
\end{equation}
Generalizing \eqref{isoSMr}-\eqref{isoSM} one finds, for isotropic fluids with $P=w\rho$,
\begin{equation}
    S\sim M^{1/(1+w)} r_0^{(d-1)w/(1+w))}\sim M^{\frac{(d-3)+(d-1)w}{(d-3) (1+w)}} 
\end{equation}
For an ordinary isotropic gas with $w=1/(d-1)$, we then have
\begin{equation}
    S\sim (A_{d-2}[M])^{(d-1)/d} \ .
\end{equation}. 
Meanwhile, for the stiffest material in any dimension $d\geq 4$,  $\rho\sim T^2$, $s\sim T$ and 
\begin{equation}
    S\sim A_{d-2}[M]
\end{equation}

If the only scales present are fundamental constants of nature (most common for traceless gases), then we can derive additional details of the scaling relations that apply even to anisotropic fluids and are functions only of $\delta$.
From  \eqref{generalgasscaling}, we have $s\propto \rho/T$ and $s\sim r^{2-\delta/2}$, which implies (retaining $c=1$)
\begin{equation}
  T\propto \frac{(\hbar G)^{(2+\delta)/4}}{Gr^{\delta/2}} \ \ ; \ \ \rho\sim \frac{T^{4/\delta}}{G^{2(\delta-1)/\delta}\hbar^{(2+\delta)/\delta}}\ \ ; \ \ s\propto \frac{1}{(\hbar G)^{(2+\delta)/4}\ r^{2-\delta/2}} 
\end{equation}
and thus, using $r_0\sim GM$,
\begin{equation}
    S \propto sr_0^3  \propto \left(\frac{GM^2}{\hbar}\right)^{(2+\delta)/4} \propto\ \left[\frac{A_2[M]}{G\hbar}\right]^{\frac12 + \delta}  \ \ . 
\end{equation}
where $A_2[M]$ is the area of a $d=4$ black hole of mass $M$. 

In general $d\geq 4$ spacetime dimensions, these relations take the form 
\begin{equation}
T\propto \frac{\hbar^q G^{q-1}}{r^{\delta/2}} \ \ ; \ \
\rho \propto  \frac{T^{4/\delta }}{G^{1+{4 (q-1)}/{\delta }} \hbar^{{4 q}/{\delta }}} \ \ ; \ \ s\propto \frac{1}{(\hbar G)^q r^{2-\delta/2}}  
\\ \ ,
\end{equation}
where
\begin{equation}
     q=1-\frac{2-\delta }{2 (d-2)} \ \ .
\end{equation}
Here $r_0^{d-3}\sim GM$, so
\begin{equation}
    S \propto s r_0^{d-3} \propto
    \left(\frac{ G^{1/(d-3)}M^{(d-2)/(d-3)}}{\hbar}\right)^q
    \propto \left[\frac{ A_{d-2}[M]}{G\hbar}\right]^q
\end{equation}
where in the last expression we are relating the entropy for a horizonless ball of gas in $d$ dimensions, with power law $\delta$ and mass $M$, to the entropy of a Schwarzschild-Tangherlini black hole with the same mass in $d$ dimensions.

\section{The Minimum Temperature of Traceless Shells}\label{app:coldestNC}

Suppose we place a series of concentric shells of traceless gas, with $\sigma=2 p$, around a black hole of mass $m$, such that the entire set has mass $M$.  We require that each such shell has non-negative mass.\footnote{For $\sigma=2p$, this is equivalent  to requiring that no shell be placed inside the photon sphere of the mass interior to it, a situation  \cite{Brady:1991np} that would cause the shell to collapse.}  As we will now show, {\it the Hawking temperature of the resulting metric is warmer than the mass of a black hole of mass $M$, except in the limit of the HBH solution where all shells have infinitesimal mass and are placed at the photon sphere of the mass interior to them.}

To prove this, take $n$ shells located at positions $r_k$ placed around a central black hole of mass $m_0=m$, where the mass parameter for $r_{k}<r<r_{k+1}$ is $m_k$. The ADM mass $M=m_n$. 
The metric near the black hole horizon is
\begin{equation}
    f(r) = f_0\left(1-\frac{2m_0}{r}\right) = f_0\ j(r)
\end{equation}
We need to show that if each shell is made of traceless gas, with $\sigma=2p$, the Hawking temperature satisfies 
\begin{equation}
    T_H = \frac{1}{8\pi m\sqrt{f_0}}\geq \frac{1}{8\pi M}\ 
\end{equation}  
[see Eq.~\eqref{THLargeR}] and thus that
\begin{equation}
    f_0 \geq \frac{m_0^2}{m_n^2}  = \frac{m^2}{M^2} \ .
\end{equation}

Continuity in $f(r)$, and the requirement that 
\begin{equation}
    f(r) \propto 1-\frac{2m_{k}}{r}  \ \ \ \ {\rm for} \ \ r_{k-1}<r<r_k
\end{equation}
implies that
\begin{equation}
    f_0 =  \prod_{k=1}^{n}
    \frac{1-\frac{2m_k}{r_k}}{1-\frac{2m_{k-1}}{r_k}} \ .
\end{equation}
We can use the fact that
\begin{equation}
   \frac{m_0^2}{m_n^2}  = \prod_{k=1}^{n} \frac{m_{k-1}^2}{m_k^2}
\end{equation}
to write $f_0\geq m^2/M^2$ as
\begin{equation}\label{neededproduct}
    \prod_{k=1}^{n}
    \frac{m_k^2\left(1-\frac{2m_k}{r_k}\right)}{m_{k-1}^2\left(1-\frac{2m_{k-1}}{r_k}\right)} \geq 1
\end{equation}

Now we appeal to the relevant Israel junction conditions \cite{Frauendiener:1990nao,Brady:1991np}.  Setting $\sigma=2p$, one can use these conditions to fix $m_k$ in terms of $m_{k-1}$, the mass interior to the $k$th shell, and $r_k$, the location of the $k$th shell.  
\begin{equation}
    m_k = r_k\frac{\frac49 r_k- m_{k-1}}{r_k-2m_{k-1}}
\end{equation}
Writing $r_k = 3m_{k-1}(1+x_k)$, and requiring  
\begin{equation}
    m_k-m_{k-1}= 2x_{k}\frac{1+2x_{k}}{1+3x_k}m_{k-1}\geq 0 \ \ \Rightarrow \ \ x_k\geq 0 \ , 
\end{equation}
we find
\begin{equation}
    \frac{m_k\sqrt{1-\frac{2m_k}{r_k}}}{m_{k-1}\sqrt{1-\frac{2m_{k-1}}{r_k}}} =  1+\frac{4x_k^3}{(1+3x_k)^2} \geq 1
\end{equation}
from which \eqref{neededproduct} follows. 

One also sees that we can only have $f_0=m^2/M^2$ if $x_k=0$ for each $k$; each shell is placed at the photon sphere.  More precisely, one takes the limit $x_k\to 0$ and  $m_k-m_{k-1}\to 0$ holding $m,M$ fixed, and obtains the smooth HBH solution.  

Thus if each $\sigma=2p$ shell is placed at or beyond the photon sphere of all previous shells, the resulting metric cannot be colder than an HBH. It thus cannot be colder than a black hole of the same ADM mass.  

A more powerful proof covering a wider range of cases and with broader physical implications is given in \cite{Riojas:2026god}.

\section{Euclidean action for an AdS-HBH in a cavity}\label{app:AdSYorkAction}

Consider an AdS$_4$ HBH with mass parameter $m$ and mass function ${\widehat{m}}$  of the usual form \eqref{massfunction0}. If the black hole is placed in a York cavity of radius $r_0$ whose boundary is at the temperature given by \eqref{AdSHBHTemplocal} for $r=r_0$, then
its Euclidean action is
\begin{eqnarray}\label{IallR}
I(r_0)&=&-\frac{\pi  r_+^2 }{ m \left(L^2 m+r_+^3\right)}\left[-\left(\frac{2}{27} L^4+ L^2 m^2- m r_+^3\right) + 2H_I(r_0,\widehat{m})\right] \ ;
     \nonumber \\ &\ & \ 
       \nonumber \\ H_I(r_0,\widehat{m}) &=& 
        \sqrt{\frac{L^2+27 m^2} {L^2+27 [\widehat{m}(r_0)]^2}} \Bigg[
        \frac{L^4}{27}+ r_0\widehat{m}(r_0)\left[r_0^2+L^2\right]
      - L^2 [\widehat{m}(r_0)]^2
      \nonumber \\ &\ & \  \ \ \ \ \ \ \ \ \ \ \ \ \ \ \ \ \ \ \ \ \ \ \ \ \ \ \   -  L^2  r_0 \widehat{m}(r_0) \sqrt{1+\frac{r_0^2}{L^2}}\sqrt{1-\frac{2 \widehat{m}(r_0)}{r_0}+\frac{r_0^2}{L^2}} \Bigg] \ .
\end{eqnarray}
One can then derive that the entropy is the same as \eqref{AdSHBHEntropy} with $M$ replaced with $\widehat{m}(r_0)$ and $\lambda_M$ replaced with
\begin{equation}
    \hat\lambda \equiv \sqrt{\frac{1}{L^2}+\frac{1}{27{\widehat{m}}(r_0)^2}} \ .
\end{equation}
One can also find our HBH result in infinite AdS$_4$ space by taking $r\to\infty$, and our flat-space results of Sec.~\ref{subsec:York} in the limit $L\to\infty$. Further results based on this formula will be given in Paper III.

\section{Christensen and Fulling for an HBH metric}
\label{app:CF2}

On a Schwarzschild black hole background, Christensen and Fulling (CF) consider \cite{Christensen:1977jc} a quantum field with a radial flow, such as one would expect for the Hawking quanta of this field.
They showed that the Bianchi identity $\nabla_\mu {\cal T}_{\mu}^{\ \nu}=0$ constrains the field's stress-tensor ${\cal T}_{\mu}^{\ \nu}$ to contain only (1)
 a constant $K$ that characterizes the luminosity of the radial flow, 
(2) the trace anomaly ${\cal T}_\alpha^{\ \alpha}(r) = g^{\alpha\beta}{\cal T}_{\alpha\beta}(r)$ and an integral thereof, 
\begin{equation}\label{Hdefn}
    H(r) = \frac12 \int_{2m}^r (r' - M) {\cal T}_\alpha^{\ \alpha}(r') > 0
\ ,
\end{equation}
and (3) the transverse pressure $P_\perp = {\cal T}_{\theta}^{\ \theta}= {\cal T}_{\phi}^{\ \phi}$ along with a second integral
\begin{equation}\label{Gdefn}
    G(r) = 2\int_{2m}^r (r' - 3M) \left[{\cal T}_\theta^{\ \theta}(r')-\frac14 {\cal T}_\alpha^{\ \alpha}(r')\right] \ .
\end{equation}
whose integrand flips sign at the photon sphere.

Only the radial flux of energy should survive into the asymptotic large-$r$ region.  Requiring consistency in this regime implies that
\begin{equation}\label{GKH}
    G(\infty) =2\int_{2m}^\infty (r' - 3M) \left[{\cal T}_\theta^{\ \theta}(r')-\frac14 {\cal T}_\alpha^{\ \alpha}(r')\right]= 2K/M^2 - H(\infty)\ .
\end{equation}
Since (for most or all fields) $K$ is positive but smaller than $H(\infty)>0$, CF conclude that $G(\infty)<0$, and so either the integrand of $G$ is negative at some values of $r>3M$ or it is positive at some values of $r<3M$.  

Applying the method of CF on the background of an HBH metric, we find a similar but stronger  conclusion, thanks to parametric separation of $m$ and $M$. Equations \eqref{Hdefn}-\eqref{Gdefn} remain true with the replacement $M\to{\widehat{m}}(r)$. For the HBH in flat space, which has $R=0$, the anomaly for a single massless scalar field takes the general form 
\begin{equation}
    \frac{1}{60\pi^2 \times 48}(C_{\mu\nu\lambda\sigma}C^{\mu\nu\lambda\sigma}+R_{\mu\nu} R^{\mu\nu}) \ .
\end{equation} 
In an HBH, from Eq.~(2.25) of Paper I, we have
\begin{eqnarray}\label{anomaly}
 (60 \pi^2) T_\alpha^\alpha =
\begin{cases}
\dfrac{ m^2}{r^6} \ ,&   r\in(r_+,\,3m),\\[3pt]
\dfrac{1}{24r^4}  \ ,&   r\in(3m,\,3M),\\[6pt]
\dfrac{ M^2}{r^6}\ ,&  r \in(3M,\,\infty).\\[8pt]
\end{cases}
\end{eqnarray}
Note the anomaly jumps discontinuously at the ocean surfaces, and is not proportional to $[{\widehat{m}}(r)]^2$.  

From this we obtain  
\begin{equation}
H(\infty) =     \frac{1}{155520\times 60\pi^2}
\left(\frac{673}{m^2}+\frac{56}{M^2}\right)
\equiv H_<(m) + H_>(M)
\end{equation}
Not only is this positive as before, forcing $G$ to be negative, its $m$ and $M$ dependence have separated.

Now assume $m$ is parametrically less than $M$.  We will also make a couple of temporary assumptions, to make our argument more transparent. First, we assume that ${\cal T}_\theta^\theta(r)$ for $r<3m$ is independent of $M$ (a reasonable assumption) and that  ${\cal T}_\theta^\theta(r)$ for $r>3M$ is independent of $m$ (less convincing.)  Notice from \eqref{anomaly} that these two facts are true of ${\cal T}_\alpha^\alpha$, however. Second, we assume that $K$, which sets the luminosity of the Hawking radiation, depends on $M$ but not $m$ (again, reasonable given that an HBH has the same temperature as a black hole of mass $M$, but not a certainty.)

We may break up the $G$ integral into three parts
\begin{eqnarray}
    G(\infty) &=&
     \left( \int_{2m}^{3m} 
    +\int_{3m}^{3M} + \int_{3M}^\infty\right)2\left[r-3{\widehat{m}}(r)\right] \left[{\cal T}_\theta^{\ \theta}(r)-\frac14 {\cal T}_\alpha^{\ \alpha}(r)\right]dr
    \nonumber \\ \nonumber \\
    &\equiv& 
     G_<(m,M) + 0 +G_>(m,M) 
\end{eqnarray}
Under our temporary assumption that the transverse pressure ${\cal T}_\theta^\theta$ is independent of $M$ for $r<3m$, and of $m$ for $r>3M$, the function $G_<$ depends only on $m$ and $G_>$ depends only on $M$.

We now apply equation \eqref{GKH}. 
The coefficient $K$ represents the radial flow at infinity, which we have assumed is $m$-independent, so equation \eqref{GKH} is a sum of terms that depend on $M$ only, plus terms ($G_<$ and $H_<$) that depend on $m$ only.  The latter give us
\begin{equation}\label{GHequal}
     G_<(m) =2 \int_{2m}^{3m} \left[r-3m\right] \left[{\cal T}_\theta^{\ \theta}(r)-\frac14 {\cal T}_\alpha^{\ \alpha}(r)\right]dr = - H_<(m)
\end{equation}
from which it follows unambiguously that ${\cal T}_\theta^{\ \theta}(r)>{\cal T}_\theta^{\ \theta}(r)-\frac14 {\cal T}_\alpha^{\ \alpha}(r)>0$ for some range of $2m<r<3m$.  
This clear conclusion cannot be drawn in CF's original argument, because the $m$ and $M$ dependence are conflated in the Schwarzschild metric.\footnote{
In fact, since ${\cal T}_\alpha^\alpha$ is known, we may constrain the transverse pressure explicitly:
\begin{equation}
    \int_{2m}^{3m} \left[3m-r\right]  {\cal T}_\theta^{\ \theta}(r)\ dr =  \frac12 H_<(m) +\frac14\int_{2m}^{3m} \left[3m-r\right] {\cal T}_\alpha^{\ \alpha}(r) dr =  \frac{241}{77760\times 60\pi^2m^2 } \ .
\end{equation}
}

We now relax our temporary assumptions. Since $H_<(m)$ diverges as $1/m^2$ in the $m\to 0$ limit, there must be other terms in \eqref{GKH} that diverge in that limit.  The Hawking flux parameter $K$ might depend on $m$, but it could only diverge if the HBH Hawking temperature diverged in that limit, which it does not.  Similarly, if $G_>(m,M)$ were to diverge in that limit, this would imply a divergent transverse pressure above the ocean as $m\to 0$, again inconsistent with what we know about the HBH Hawking temperature, which remains finite in that limit.  The only other term that can go as $1/m^2$ is $G_<(m,M)$, and setting the diverging terms equal gives \eqref{GHequal}.

This transverse pressure appearing in \eqref{GHequal} is per scalar quantum field; in a theory with multiple fields on an HBH background, this pressure will be larger.  Nevertheless, it is parametrically of order $\hbar/r^4\sim \hbar/m^4$ below the ocean, with a numerically small coefficient. As such it is far less than the ``classical'' ocean pressure of \eqref{gassmassfunction}, $P_\perp= 1/24\pi Gr^2$, which is independent of the number and type of fields in the theory. 

\end{document}